\documentclass[aps,prb,reprint,superscriptaddress,amsmath,amssymb,showpacs]{revtex4-2}

\usepackage{amsfonts,amsbsy,amsthm}
\usepackage{xcolor}
\usepackage{bm}
\usepackage{graphicx}
\usepackage{yhmath}
\usepackage{slashed}
\usepackage{mathtools}

\def\XXint#1#2#3{{\setbox0=\hbox{$#1{#2#3}{\int}$ }
\vcenter{\hbox{$#2#3$ }}\kern-.6\wd0}}

\newcommand{\B}{{\bf B}}
\newcommand{\E}{{\bf E}}
\newcommand{\cE}{{\mathcal{\bm E}}}
\newcommand{\J}{{\bf J}}

\newcommand{\K}{{\bf K}}

\newcommand{\m}{{\bf m}}
\newcommand{\e}{{\bf e}}

\newcommand{\p}{{\bf p}}
\renewcommand{\j}{{\bf j}}
\newcommand{\cJ}{{\mathcal{\bm J}}}

\newcommand{\q}{{\bm{q}}}

\newcommand{\bx}{\boldsymbol x}
\newcommand{\bD}{\boldsymbol{D}}
\newcommand{\bd}{\boldsymbol{d}}
\newcommand{\px}{\partial}
\newcommand{\bpx}{\bar{\partial}}
\newcommand{\bq}{\boldsymbol q}

\newcommand{\taub}{\boldsymbol{\tau}}

\newcommand{\tvpb}{\boldsymbol{V}}

\newcommand\Rep{\textrm{Re}}
\newcommand\Imp{\textrm{Im}}

\newcommand{\bsigma}{\boldsymbol{\sigma}}

\definecolor{bvio}{rgb}{0.54, 0.17, 0.89}

\definecolor{fgr}{rgb}{0.13, 0.55, 0.13}

\begin{document}

\title{Optical response of alternating twisted trilayer graphene}

\author{Dionisios Margetis}
\email{diom@umd.edu}

\affiliation{Department of Mathematics, and Institute for Physical Science
and Technology, University of Maryland, College Park, Maryland 20742, USA}

\author{Guillermo G\'omez-Santos}
\email{guillermo.gomez@uam.es}

\affiliation{Departamento de F\'{\i}sica  de la Materia Condensada,
Instituto Nicolás Cabrera and Condensed Matter Physics Center (IFIMAC),
Universidad Autónoma de Madrid, E-28049 Madrid, Spain}

\author{Tobias Stauber}
\email{tobias.stauber@csic.es}

\affiliation{Instituto de Ciencia de Materiales de Madrid, CSIC, E-28049 Madrid, Spain}

\date{\today}

\begin{abstract}
We study the optical response of the alternating twisted trilayer graphene by making use of a unitary transformation for the trilayer Hamiltonian and the Kubo formulation of linear response theory. The layer-resolved optical conductivities are expressed in terms of contributions from {\em effective} twisted bilayer and single-layer systems along with their coupling. We show that the in-plane magnetic response is proportional to this coupling between the twisted bilayer and single-layer systems; and, due to the different energy scales, the in-plane magnetic response is negligibly small. We also formulate a {\em local} electro-magnetic response that involves the vertical gradients of the magnetic field and moment. 
\end{abstract}

\maketitle

\section{Introduction}
\label{sec:Intro}
In 2018, it was shown that the twist angle between two graphene layers can be tuned so that correlated insulator phases~\cite{Cao_2018} and superconductivity~\cite{Cao_2018unconv} emerge for small carrier densities around charge neutrality. Other interesting phases such as anomalous Hall ferromagnetism~\cite{Sharpe19,Sharpe21} or fractional Chern numbers have also been observed~\cite{Ledwith20,Xie2021,Pierce2021}. These phenomena are related to the flat bands around charge neutrality that have been predicted theoretically~\cite{suarezmorell,bistritzer}. 

Since then, the study of twisted geometries of van der Waals heterostructures, often referred to as moir\'e materials~\cite{Andrei20}, has become an active area of research. The subjects of these investigations include twisted bilayers composed of various transition metal dichalcogenides~\cite{Jin19,Shimazaki20,Wang20,Wang22,Park23,Park23b,Wu23,Jiabin23}, besides the twisted bilayer graphene. Moreover, the number of sheets in moir\'e graphene systems has been increased to up to five alternately twisted layers~\cite{Park22,Ledwith22}. Recently, also quasi-three-dimensional twisted structures were fabricated, which exhibit an intrinsic enhanced  chirality~\cite{Mannix22,Ji24}.

Lately, special emphasis has been placed on the study of the twisted trilayer graphene~\cite{Park21,Zeyu21,Cao21,Carr20,Zhu20,Ramires21,Phong21,Zhang21,FangXie21b,Fischer21,Christos22,Turkel22,LiuXiaoxue22,Lin22,Christos23,Gonzalez23,Devakul23,Kim23,Ren24}. 
This is the minimal system that can form not only quasi-commensurate but also incommensurate structures. Quasi-commensurate structures can be described by two moir\'e vectors, whereas incommensurate structures are usually described by four moir\'e vectors - two for each of the moir\'e lattices that are formed by the two pairs of consecutive layers (say, layers 1 and 2, and layers 2 and 3, respectively)~\cite{Zhu20,Becker20,Watson22,Nakatsuji23,Popov23}. Interestingly, superconductivity has been found in both types of structures, that is, commensurate~\cite{Park21,Zeyu21,Cao21,Kim23} and incommensurate trilayer structures~\cite{Uri23,Xia24}.

Notably, optical experiments on moir\'e systems have also attracted considerable attention~\cite{Kim16,Ma22,Yu22,Moreno24}. A goal is to control the phase, amplitude or polarization state of light from the response of multilayered structures. 

In this paper, we analytically study the electromagnetic response of the twisted trilayer graphene.
Our approach is based on symmetry considerations and microscopic principles of linear response theory. We focus on the alternating-twist configuration, in which $\theta_{12}=-\theta_{23}$ where $\theta_{ij}$ denotes the angle between layers $i$ and $j$. In this case, in the presence of mirror symmetry with respect to the central layer, also preserved after including in-plane and out-of-plane relaxation~\cite{Ceferino24}, we show that the in-plane magnetic response is negligibly small. By mirror symmetry, there is no optical activity; however, we formulate a local magneto-electric (chiral) coupling that involves vertical gradients of the magnetic field and moment. 

In our analysis, we make use of a unitary transformation, introduced in Ref.~\onlinecite{Khalaf19}, that converts the trilayer system into a combination of effective twisted bilayer and single-layer systems. Linear response theory for the trilayer eventually couples these two systems. We derive a formula for this coupling, and estimate its magnitude. We also describe, in terms of contributions from these two systems, the layer-resolved conductivities that enter the optical response. In particular, we analytically show how the in-plane magnetic response arises from the coupling of the two effective systems. 

The alternating-twist trilayer system under mirror symmetry investigated here is one of the most stable layered structures, minimizing the configuration stacking fault energy. This remarkable mechanical stability renders this system prototypical, and is another motivation for our work. Furthermore, in appendices  we extend the analysis for the optical conductivity to alternating-twist trilayers in which the mirror symmetry with respect to the middle layer is broken because of the different interlayer tunneling. 

A highlight of our results is the connection of the optical response of the trilayer system to microscopic parameters of effective Hamiltonians for simpler systems. Thus, we are able to illustrate the important role of the large moir\'e length. For example, the effect of this scale renders negligible the relative contribution of the in-plane magnetic response. In this paper, we focus on analytical predictions that explicitly reveal the interplay of two scales in the optical response. Hence, we choose not to use or rely on numerical simulations in this work. 

 
The remainder of the paper is organized as follows. Section~\ref{sec:micro} focuses on the conductivity tensor via the Kubo formulation. In particular, we discuss the roles of the effective twisted bilayer and single-layer systems and the coupling term between them. In Sec.~\ref{sec:MagneticMoment}, we formulate the optical response in terms of an in-plane electric moment and the vertical gradient of the in-plane magnetic moment, which are coupled. In Sec.~\ref{subsec:eff-parameters}, we discuss the in-plane magnetic response by giving an analytical expression and estimates. In Sec.~\ref{sec:conclusion}, we conclude the paper with a summary of our main results. The appendices mainly provide technical derivations relevant to the main text. 

Throughout the paper, boldface symbols denote vectors or matrices that pertain to the ($x$- and $y$-) directions of the reference plane. The $e^{-i\omega t}$ time dependence is assumed  ($\omega$ is the frequency).

\section{Microscopic theory for layer-resolved conductivities}
\label{sec:micro}
In this section, we provide key ingredients and results of the Kubo formulation for the trilayer system. 
The general layer-resolved response theory in the frequency domain for a system with $n$ layers is given by~\cite{Stauber18,Stauber18b}  
\begin{align}\label{eq:J-response}
\J_\ell=\sum_{\ell'=1}^n\bsigma^{\ell\ell'}\E_{\ell'}\quad (\ell=1,\,\ldots,\,n)\;,
\end{align}
where $\J_\ell$ and $\E_\ell$ denote the macroscopic surface current density and electric field in layer $\ell$, respectively. The $2\times2$ matrices $\bsigma^{\ell\ell'}(\omega)$ have elements defined by $\sigma^{\ell\ell'}_{\nu\nu'}(\omega)=i\frac{e^2}{\omega+i\delta}\chi^{\ell\ell'}_{\nu\nu'}(\omega+i\delta)$, as $\delta\downarrow 0$, with the current-current response function
\begin{align}\label{eq:resp-func}
\chi^{\ell\ell'}_{\nu\nu'}(\omega)=-\frac{i}{\hbar}\int_0^\infty dt\ e^{i\omega t}\langle[ j^\ell_\nu(t),j^{\ell'}_{\nu'}(0)]\rangle\;,
\end{align}
where $j^\ell_\nu(t)$ is the $\nu$-directed current operator ($\nu=x,y$) in layer $\ell$ in the interaction picture, and $\langle \cdot \rangle$ is the equilibrium average.  

\begin{figure}[h]
\includegraphics[scale=0.34,trim=0in 0.8in 0in 0.4in]{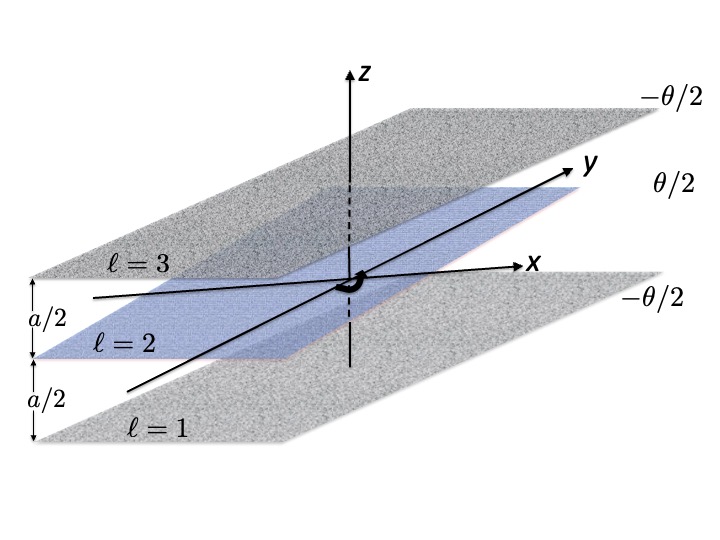}
\caption{Schematic of the alternating-twist trilayer configuration.
Three infinite flat graphene sheets labeled by $\ell=1,\,2,\,3$ are parallel to the $xy$-plane, and have twist angles $(-1)^{\ell}\theta/2$ and interlayer distance equal to $a/2$ ($0<\theta< \pi/2$). The layers are immersed in a homogeneous medium.}
\label{fig:Geometry}
\end{figure}

Here, we consider a moir\'e multilayer with $n=3$. The geometry of the alternating-twist trilayer system is shown in Fig.~\ref{fig:Geometry}. The twist angle of layer $\ell$ is $\theta_\ell=(-1)^{\ell}\theta/2$ where $0<\theta<  \pi/2$ and $\ell=1,\,2,\,3$; the interlayer distance is $a/2$. Furthermore, we mainly consider a mirror-symmetric configuration. Hence, the matrices $\bsigma^{\ell\ell'}$ satisfy $\bsigma^{11}=\bsigma^{33}$, $\bsigma^{12}=\bsigma^{32}$, and $\bsigma^{21}=\bsigma^{23}$. 

Let us recall that the symmetry group of the twisted bilayer system is $D_3$~\cite{Stauber23}. However, by the mirror symmetry of the alternating-twist trilayer, the symmetry group now is $D_{3h}$. This additional symmetry doubles the number of representations, i.e., we have four one-dimensional and two two-dimensional representations in total. 

Due to the rotational (three-fold) symmetry of the system, we can write each $2\times2$ conductivity matrix as follows~\cite{Stauber18}:
\begin{align}\label{eq:sigma-ll'}
\bsigma^{\ell\ell'}=\sigma^{\ell\ell'}_0{\bf 1}+i\sigma^{\ell\ell'}_{xy}{\boldsymbol \tau}_y\;,
\end{align}
where ${\boldsymbol \tau}_y$ denotes the {\it y}-Pauli matrix.
Because of the mirror symmetry of the trilayer configuration, we only have four independent response functions, namely, $\sigma^{11}_0$, $\sigma^{12}_0$, $\sigma^{13}_0$ and $\sigma^{22}_0$. These are related to the four one-dimensional representations of the group $D_{3h}$. 

Moreover, time reversal symmetry implies $\sigma^{\ell\ell'}_{\nu\nu'}=\sigma^{\ell'\ell}_{\nu'\nu}$; see Appendix~\ref{app:time-reverse}. Because of time reversal and mirror symmetry, we may further include the parameters $\sigma^{12}_{xy}$ and $\sigma^{13}_{xy}$, which are related to the two two-dimensional representations of $D_{3h}$. These parameters can give rise to chiral phenomena such as circular dichroism~\cite{Barron,Barron04,Kim16,Moreno24} and layer-resolved Hall physics in the d.c. limit~\cite{Stauber20b,Bahamon24}. Here, $\sigma^{13}_{xy}=0$, since we consider the $\q=0$-response and the first and third layer have the same orientation; see Fig.~\ref{fig:Geometry}. The parameter $\sigma^{12}_{xy}$ will be kept finite; however, no chiral response can emerge due to the inherent mirror symmetry. Still, we can formulate constitutive equations  that involve $\sigma^{12}_{xy}$ and lead to {\em local} chirality; see  Sec.~\ref{sec:MagneticMoment}.

\subsection{Kubo formalism}
\label{subsec:formalism-cond}

By  Eq.~\eqref{eq:resp-func}, the surface conductivity matrix elements in the frequency domain are given by
\begin{align}
	\sigma_{\nu\nu'}^{\ell\ell'}(\omega)&=\frac{e^2}{\hbar(\omega+i\delta)}\sum_{\{\vert \psi\rangle\}}f(E_\psi)\notag\\
	& \times \int_0^\infty dt\ e^{i(\omega+i\delta)t}\langle \psi \vert [j_\nu^{\ell}(t), j_{\nu'}^{\ell'}(0)]\vert \psi\rangle\;, \label{eq:Kubo-formula-freq}
\end{align}	
as $\delta\downarrow 0$, for real frequencies $\omega$ ($\ell,\,\ell'=1,\,2,\,3$ and $\nu,\,\nu'=x,\,y$).
Here,  $\vert\psi \rangle$ is any normalized eigenvector of the trilayer Hamiltonian with eigenvalue $E_\psi$; and $f(E)$ is the Fermi-Dirac distribution. Regarding the $K$-valley, we use the Hamiltonian $\mathcal H^{tri}_K$ of Appendix~\ref{app:Ham-3layer}; and symmetrize the ensuing conductivity because of time reversal symmetry (Appendix~\ref{app:time-reverse}). In Eq.~\eqref{eq:Kubo-formula-freq}, $j_\nu^{\ell}(t)$ has dimensions of inverse time, or frequency. 

To describe the $K$-valley term $\sigma_{\nu\nu',K}^{\ell\ell'}$, we rewrite Eq.~\eqref{eq:Kubo-formula-freq} by unitarily transforming each eigenvector 
$\vert\psi\rangle$, while $\mathcal H^{tri}_K$ is transformed into the direct sum of the single-layer and effective twisted bilayer Hamiltonians $\mathcal H_1$ and $\mathcal H_2$, respectively;  see Eqs.~\eqref{eq:eff-H1-def} and~\eqref{eq:eff-H2-def} in Appendix~\ref{app:Ham-3layer}. The transformed eigenvectors of  $\mathcal H^{tri}_K$ are $\vert \underline{\psi}\rangle=\mathfrak S^\dagger\vert\psi\rangle$, where $\mathfrak S$ is defined in Eq.~\eqref{eq:T-matrix} or~\eqref{eq:T-matrix-mirror} (Appendix~\ref{app:Ham-3layer}); and are  related to either the effective twisted bilayer or single-layer Hamiltonian:
\begin{align}\label{eq:G-transf-vector}  
	\vert \underline{\psi}\rangle_b=\begin{pmatrix}
	\vert b \rangle_1 \\
	\vert b\rangle_2 \\
	 0  \\
	 \end{pmatrix}\ \ \mbox{,}\ \ 	 
	 \vert \underline{\psi}\rangle_s=\begin{pmatrix}
	0 \\
	 0 \\
	 \vert s \rangle \\
	 \end{pmatrix}\;,
	\end{align}
where $(\vert  b \rangle_1, \vert b\rangle_2)^T$ and $\vert s \rangle$ denote any 
normalized-to-unity eigenvector of the effective bilayer Hamiltonian
$\mathcal H_2$ and monolayer Hamiltonian $\mathcal H_1$, respectively. Each of the  indices $b$ and $s$ amounts to the combined band index and quasi-momentum of the moir\'e Brillouin zone; and $\langle \bx\vert\underline{\psi}\rangle$ consists of Bloch functions. We thus have
\begin{equation*}
\mathcal H_{2} \begin{pmatrix}
 \vert b\rangle_1 \\
 \vert b \rangle_2 \\	
 \end{pmatrix}
 = E_{b} 
 \begin{pmatrix}
\vert b\rangle_1\\
\vert b\rangle_2
\end{pmatrix}\;,\quad 
	\mathcal H_{1} 
 \vert s \rangle \\	
 = E_{s} 
\vert s\rangle\;.
\end{equation*}
By this notation, Eq.~\eqref{eq:Kubo-formula-freq} becomes
\begin{widetext}
\begin{align}
	&\sigma_{\nu\nu',K}^{\ell\ell'}(\omega)=\frac{e^2}{\hbar(\omega+i\delta)}\int_0^\infty\!\!\!\! dte^{i(\omega+i\delta)t}\Big \{\sum_{b}f(E_b){}_b\langle \underline{\psi}\vert[\underline{j}_{\nu,K}^{\ell}(t), \underline{j}_{\nu',K}^{\ell'}(0)]\vert \underline{\psi}\rangle_b 
	 + \sum_{s}f(E_s) {}_s\langle \underline{\psi}\vert[\underline{j}_{\nu,K}^{\ell}(t), \underline{j}_{\nu',K}^{\ell'}(0)]\vert \underline{\psi}\rangle_s \Big\}\;,
	 	 \label{eq:Kubo-formula-freq-transf}
\end{align}
\end{widetext}	
where $\underline{j}^{\ell}_{\nu,K}=\mathfrak S^\dagger j^{\ell}_{\nu,K} \mathfrak S$. For details, see Appendix~\ref{app:conductivity}.

\subsection{Transforming conductivity tensor via mirror symmetry}
\label{subsec:formalism-cond}

Next, we focus on the mirror-symmetric setting, where the electron tunneling for layers $1,\,2$ is the same as that for layers $2,\,3$. In Appendix~\ref{app:conductivity}, we discuss the more general setting without mirror symmetry by keeping the alternating twists, using $\theta_{12}=-\theta_{23}$. 

Let $\bsigma_{BL}^{\varsigma\varsigma'}$ ($\varsigma, \varsigma'=1,\,2$) and $\bsigma_{SL}$ denote the $2\times 2$ conductivity block matrices of the effective twisted bilayer and single-layer systems. These quantities have been discussed in previous work~\cite{Stauber18}, and can be considered as known. In Appendix \ref{app:conductivity}, they are expressed in terms of matrix elements of pseudo-spin wave functions. 

The connection between the trilayer response and $\bsigma_{BL}^{\varsigma\varsigma'}$ and $\bsigma_{SL}$ can be obtained by applying to the currents the following transformation:
\begin{align}\label{eq:J-BL-SL}
\begin{pmatrix}
\tilde\J_1 \\
\tilde \J_2 \\
\tilde \J_3 	
\end{pmatrix}&=
\begin{pmatrix}
\tfrac{1}{\sqrt{2}} & 0 & \tfrac{1}{\sqrt{2}} \\
0    &   1    &  0 \\
\tfrac{1}{\sqrt{2}} & 0 & -\tfrac{1}{\sqrt{2}}	
\end{pmatrix}
\begin{pmatrix}
\J_1 \\
\J_2 \\
\J_3 	
\end{pmatrix}\;.
\end{align}
The electric fields are transformed by the same matrix into ($\tilde\E_1$, $\tilde\E_2$, $\tilde \E_3$). The transformation for the current densities here follows directly from the conversion of the trilayer Hamiltonian into a direct sum of effective Hamiltonians  (Appendix~\ref{app:Ham-3layer}). By the microscopic analysis of Appendix~\ref{app:conductivity}, results of which we show below, the layer-resolved response implies a relation of the form
\begin{align}\label{eq:J-E-effective}
\begin{pmatrix}
\tilde\J_1 \\
\tilde \J_2 \\
\tilde \J_3 	
\end{pmatrix}=	\frac{1}{2}\begin{pmatrix}
(\bsigma_{BL}^{11}+\bsigma_{SL})& \sqrt{2}\bsigma_{BL}^{12} & 0\\
\sqrt{2}\bsigma_{BL}^{21} & 2\bsigma_{BL}^{22} & 0\\
0  & 0 & \bsigma_c\	
\end{pmatrix}
			\begin{pmatrix}
\tilde\E_1 \\
\tilde \E_2 \\
\tilde \E_3 	
\end{pmatrix}\;,
\end{align}
where 
\begin{align*}
\bsigma_{SL}&=\sigma_e^{(1)}{\bf 1}\;,\;
	\bsigma_{BL}^{11}=\bsigma_{BL}^{22}=\sigma_{e}^{(2)}{\bf 1}\;,\;\bsigma_c=\sigma_{c}{\bf 1}\;,\\
\bsigma_{BL}^{12}&=(\bsigma_{BL}^{21})^T=\sigma_{e}^{12}{\bf 1}+i\sigma_{e,xy}^{12}{\boldsymbol \tau}_y\;.
\end{align*}
Here, we introduced the scalar functions $\sigma_e^{(\varsigma)}(\omega)$, $\sigma_{e}^{12}(\omega)$, $\sigma_{e,xy}^{12}(\omega)$, and $\sigma_{c}(\omega)$, to be described in Sec.~\ref{subsec:explicit_cond}. The above matrix forms account for time reversal symmetry and isotropy (see Appendix~\ref{app:conductivity}).

By transforming Eq.~\eqref{eq:J-E-effective} back to the layer-resolved response of Eq.~\eqref{eq:J-response}, we identify the $\bsigma^{\ell\ell'}$ matrices as follows ($\ell,\,\ell'=1,\,2,\,3$):
\begin{align}\label{eq:sigma11-mirror}
	\bsigma^{11}&=\tfrac{1}{4}(\bsigma_{BL}^{11}+\bsigma_{SL}+\bsigma_c)\,,\\
\label{eq:sigma13-31-mirror}
	\bsigma^{13}&=\tfrac{1}{4}
	(\bsigma_{BL}^{11}+\bsigma_{SL}-\bsigma_c)\,,\\
	\label{eq:sigma12-mirror}
	\bsigma^{12}&=\tfrac{1}{2}\bsigma^{12}_{BL}\,,\,\bsigma^{22}=\bsigma_{BL}^{22}\;.
\end{align}
The other components are obtained by mirror and time reversal symmetry; cf. Eq.~\eqref{eq:sigma-ll'}.

From the above equations, the optical responses of Eq. (\ref{eq:J-response}) can thus be expressed in terms of $\sigma_e^{(1,2)}$, $\sigma_{e(,xy)}^{12}$ and $\sigma_{c}$. Except for $\sigma_{c}$, these conductivities are known, in principle, from the electromagnetic responses of the isolated single- and twisted bilayer graphene~\cite{Stauber18}.  In contrast, $\sigma_{c}$ stands out because, even though the trilayer Hamiltonian decouples into effective single-layer and bilayer systems, the currents involved in $\sigma_{c}$ mix both of these sub-systems.

\subsection{Explicit expressions for conductivities}
\label{subsec:explicit_cond}
We now describe the parameters of $\bsigma_{BL}^{\varsigma\varsigma'}$, $\bsigma_{SL}$ and $\bsigma_c$ via inner products of pseudo-spin wave functions. For notational convenience, let ${\boldsymbol \tau}_{\pm}=\tfrac{1}{2}({\boldsymbol \tau}_x\pm i{\boldsymbol \tau}_y)$ where $\tau_{x,y}$ are the $x,y$-components of the Pauli matrices. 

First, the matrix $\bsigma_c(\omega)$ emerges from the overlap of Bloch eigenvectors of the two effective systems in the Kubo trace formula.  By $\bsigma_c=\mathrm{diag}(\sigma_{c}, \sigma_{c})$, we derive
\begin{align}\label{eq:sigma-coupling}
	\sigma_{c}&=-i 2g_vg_s \frac{v_F^2}{A_{m}}\frac{e^2}{\hbar(\omega+i\delta)}\sum_{bs}\frac{(f_b-f_s)\omega_{bs}}{(\omega+i\delta)^2-\omega_{bs}^2}\notag\\
&\qquad \times \left(|{}_1\!\langle b\vert \tau_+\vert s\rangle|^2+|{}_1\!\langle b\vert \tau_-\vert s\rangle|^2\right)\;,
\end{align}
where $\omega_{bs}=(E_b-E_s)/\hbar$, $f_{b(s)}=f(E_{b(s)})$, $g_v$ ($g_s$) is the valley (spin) degeneracy factor, $v_F$ is the Fermi velocity of monolayer graphene, and $A_m$ is the area of the reference moir\'e cell; see Appendix~\ref{app:conductivity}. This expression is simplified in view of $|{}_1\!\langle b\vert \tau_+\vert s\rangle|^2=|{}_1\!\langle b\vert \tau_-\vert s\rangle|^2$, because of a symmetry of the effective Hamiltonians $\mathcal H_{1,2}$ under the interchange of the sublattice indices. The relative magnitude of $\sigma_{c}$ is estimated in Sec.~\ref{subsec:eff-parameters}.

We will now discuss the response functions, also present in the bilayer system~\cite{Stauber18}.
The first element is the chirality parameter, $\sigma_{xy}^{12}$, from the effective bilayer system. This equals 
\begin{align}\label{eq:chiral-mirror}
&\sigma_{xy}^{12}(\omega)=\tfrac{1}{2}\sigma_{e,xy}^{12}= i g_v g_s \frac{v_F^2}{A_m} \frac{e^2}{\hbar(\omega+i\delta)} \sum_{bb'}(f_b-f_{b'}) \notag\\
& \times \frac{\omega_{bb'} \,\Imp\bigl\{e^{-i\theta} 
\bigl(
{}_2\langle b'\vert \tau_- \vert b\rangle_2\bigr)  
\bigl({}_1\!\langle b\vert \tau_+ \vert b' \rangle_1\bigr)\bigr\}}{(\omega+i\delta)^2-\omega_{bb'}^2}
\end{align}
where $\omega_{bb'}=(E_b-E_{b'})/\hbar$. For $\theta=0$, $\sigma_{xy}^{12}$ must vanish. This property is derived by noting that for $\theta=0$ we have  
${}_1\!\langle b\vert \tau_{\pm} \vert b'\rangle_1= {}_2\langle b\vert \tau_{\pm} \vert b'\rangle_2$, switching layers in $\mathcal H_2$, where ${}_{1(2)}\!\langle b\vert \tau_{\pm} \vert b'\rangle_{1(2)}=\bigl({}_{1(2)}\!\langle b'\vert \tau_{\mp} \vert b\rangle_{1(2)}\bigr)^*$. More generally, we have $\sigma_{xy}^{12}(\omega;\theta)=-\sigma_{xy}^{12}(\omega; -\theta)$; see also Ref.~\onlinecite{Stauber18}.

The remaining response parameters are given by 
\begin{align}
\sigma_e^{(1)}&=-i 2g_v g_s \frac{v_F^2}{A_m}\frac{e^2}{\hbar(\omega+i\delta)} \sum_{ss'}\frac{(f_s-f_{s'})\omega_{ss'}}{(\omega+i\delta)^2-\omega_{ss'}^2}\notag\\
&\qquad \times |\langle s\vert \tau_+\vert s'\rangle|^2\,,
\end{align}
\begin{align}\label{eq:sigmae-2}
	\sigma_e^{(2)}&=-i g_v g_s\frac{v_F^2}{A_m}\frac{e^2}{\hbar (\omega+i\delta)}\sum_{bb'}\frac{(f_b-f_{b'})\omega_{bb'}}{(\omega+i\delta)^2-\omega_{bb'}^2}\notag\\
	& \qquad \times \bigl(|{}_1\!\langle b\vert \tau_+ \vert b'\rangle_1|^2+ |{}_2\langle b'\vert \tau_- \vert b\rangle_2|^2\bigr)\,,
\end{align}
\begin{align}\label{eq:sigmae-12}
\sigma_e^{12}&= -i 2g_vg_s\frac{v_F^2}{A_m} \frac{e^2}{\hbar(\omega+i\delta)} \sum_{bb'}\frac{(f_b-f_{b'})\omega_{bb'}}{(\omega+i\delta)^2-\omega_{bb'}^2}\notag\\
&\qquad \times  \Rep\bigl\{e^{-i\theta}
\bigl({}_1\!\langle b\vert \tau_+\vert b'\rangle_1\bigr) \bigl({}_2\langle b'\vert \tau_- \vert b\rangle_2\bigr)\bigr\}\;.
\end{align}
By Eqs.~\eqref{eq:sigmae-2} and~\eqref{eq:sigmae-12}, regarding the effective bilayer system, we can show that $\sigma_e^{\eta}(\omega;\theta)= \sigma_e^{\eta}(\omega;-\theta)$; see also Ref.~\onlinecite{Stauber18}.

\section{Coupling between electric and local magnetic fields}
\label{sec:MagneticMoment}
In the preceding analysis, we have formulated the linear response in terms of conductivities. However, the emergence of chiral effects is often described in terms of a coupled response to electric and magnetic fields.

For even modes, the trilayer responses can be read off from the responses of the single-layer graphene and twisted bilayer graphene, which are already known; see Eq.~\eqref{eq:J-E-effective}. Given the chiral nature of twisted bilayer graphene, and to make contact with Ref.~\onlinecite{Stauber18}, we will also employ a magnetic language. Notice, though, that the mirror-symmetric trilayer does not exhibit optical activity. Therefore, even though some equations for the trilayer will look formally very similar to those of the twisted bilayer graphene~\cite{Stauber18},  magnetic moments and magnetic fields will enter through layer differences, considered as ``gradients'' in the $z$ direction, to comply with space-inversion symmetry \cite{Barron04}.

\subsection{{Modes} under mirror symmetry}
We define the following transformation of the original electric fields that will relate the current response to effective electric and magnetic dipoles: 
\begin{align}
\label{eq:NormalModes}
\begin{pmatrix}
\boldsymbol \cE_+^+ \\
\boldsymbol \cE_+^- \\ 
\boldsymbol \cE_-
\end{pmatrix}=\boldsymbol M                                                                                      \begin{pmatrix} \E_1 \\ \E_2\\ \E_3\end{pmatrix}\;,
\end{align}
with 
\begin{align}\label{eq:matrix-M}
\boldsymbol M=
\begin{pmatrix} 
\sqrt{1/3}&\sqrt{1/3}& \sqrt{1/3}\\
                                                      \sqrt{1/6}& -\sqrt{2/3} & \sqrt{1/6}\\
                                                        \sqrt{1/2}&0&-\sqrt{1/2}
\end{pmatrix}\;. 
\end{align}    
In the above matrix, each entry expresses a $2 \times 2$ block matrix. Up to normalizing constants, this matrix can be understood as follows: The first row is the layer-averaged field, the second row is the (layer-discrete) second derivative of the field, and the third row is the (layer-discrete) first derivative of the field.                                                                             

The original current densities are transformed in the same fashion as the original electric fields (Eq.~\eqref{eq:NormalModes}), according to the equation
\begin{align}\label{eq:transformed-currents}
\begin{pmatrix}
\boldsymbol \cJ_+^+ \\
\boldsymbol \cJ_+^- \\ 
\boldsymbol \cJ_-
\end{pmatrix}=\boldsymbol M  
\begin{pmatrix}
\boldsymbol \J_1 \\
\boldsymbol \J_2 \\ 
\boldsymbol \J_3
\end{pmatrix}\;.
\end{align}
 
Since the total system is mirror symmetric, the odd mode, $\boldsymbol \cJ_-$, is again decoupled from the two even modes. Thus, we obtain the following relations:
\begin{align}
\label{eq:CoupledE}
\begin{rcases*}
\begin{aligned}
\boldsymbol \cJ^+_+&=\sigma_+^+ \boldsymbol \cE^+_++\sqrt{2}\sigma_+^{+-} \boldsymbol \cE^-_++\sqrt{2}\sigma_{xy}^{12}(\e_z\times \boldsymbol \cE_+^-)\;,\\
\boldsymbol \cJ_+^-&=\sigma_+^- \boldsymbol \cE_+^-+\sqrt{2}\sigma_+^{+-} \boldsymbol \cE^+_+-\sqrt{2}\sigma_{xy}^{12}(\e_z\times \boldsymbol \cE_+^+)\;,\\
\boldsymbol \cJ_-&=(\sigma_{11}^0-\sigma_{13}^0)\boldsymbol \cE_-\;.
\end{aligned}
\end{rcases*}
\end{align}
Here, $\mathbf{e}_z$ is the $z$-directed Cartesian unit vector, and 
\begin{align}
\label{eq:sigma+-}
\begin{rcases*}
\begin{aligned}	
3\sigma_+^+&=2(\sigma_0^{11}+\sigma_0^{13})+\sigma_0^{22}+4\sigma_0^{12}\;,\\
3\sigma_+^-&=(\sigma_0^{11}+\sigma_0^{13})+2\sigma_0^{22}-4\sigma_0^{12}\;,\\
3\sigma_+^{+-}&=(\sigma_0^{11}+\sigma_0^{13})-\sigma_0^{22}-\sigma_0^{12}\;.
\end{aligned}
\end{rcases*}
\end{align}
Recall that by Eqs.~\eqref{eq:sigma11-mirror}--\eqref{eq:sigma12-mirror} the conductivities $\sigma_0^{\ell\ell'}$ are expressed in terms of the effective single-layer and twisted bilayer responses. Furthermore, we have $\sigma_c=2(\sigma_{11}^0-\sigma_{13}^0)$ from Eqs.~\eqref{eq:sigma11-mirror} and \eqref{eq:sigma13-31-mirror}.

\subsection{Effective local description for  modes}
\label{subsec:in-plane-magn}
Let us now focus on the  modes that define the response of the trilayer. We  can formulate the constitutive equations with respect to electric and local magnetic fields~\cite{Stauber18}. From the discrete (layer-resolved) version of Maxwell's equations, we write
\begin{align}
\E_\parallel&={(\E_1+\E_2+\E_3)/3}\;,\label{eq:E-par-sum}\\
i\omega(a/2)\B_\parallel^{12}&=\e_z\times(\boldsymbol \E_2-\boldsymbol\E_1)\;,\label{eq:B-par-12}\\
i\omega(a/2)\B_\parallel^{23}&=\e_z\times(\boldsymbol \E_3-\boldsymbol\E_2)\;. \label{eq:B-par-23}
\end{align}
{Equation~\eqref{eq:E-par-sum} is the average electric filed.} Equations~\eqref{eq:B-par-12} and~\eqref{eq:B-par-23} are discrete versions of the Maxwell-Faraday law (or, the third Maxwell equation).  Therefore, $\B_\parallel^{\ell\ell'}$ is the (average) magnetic field between layers $\ell$ and $\ell'$. For mirror-symmetric modes, when $\E_1=\E_3$, there is no net magnetic field, i.e., $\B_\parallel^{12}+\B_\parallel^{23}=0$. Nevertheless, a finite value of $\B_\parallel^{12}$ ($\B_\parallel^{23}$) can be viewed as a vertical gradient of the magnetic field, as previously noted.

\begin{figure}[h]
\includegraphics[scale=0.25]{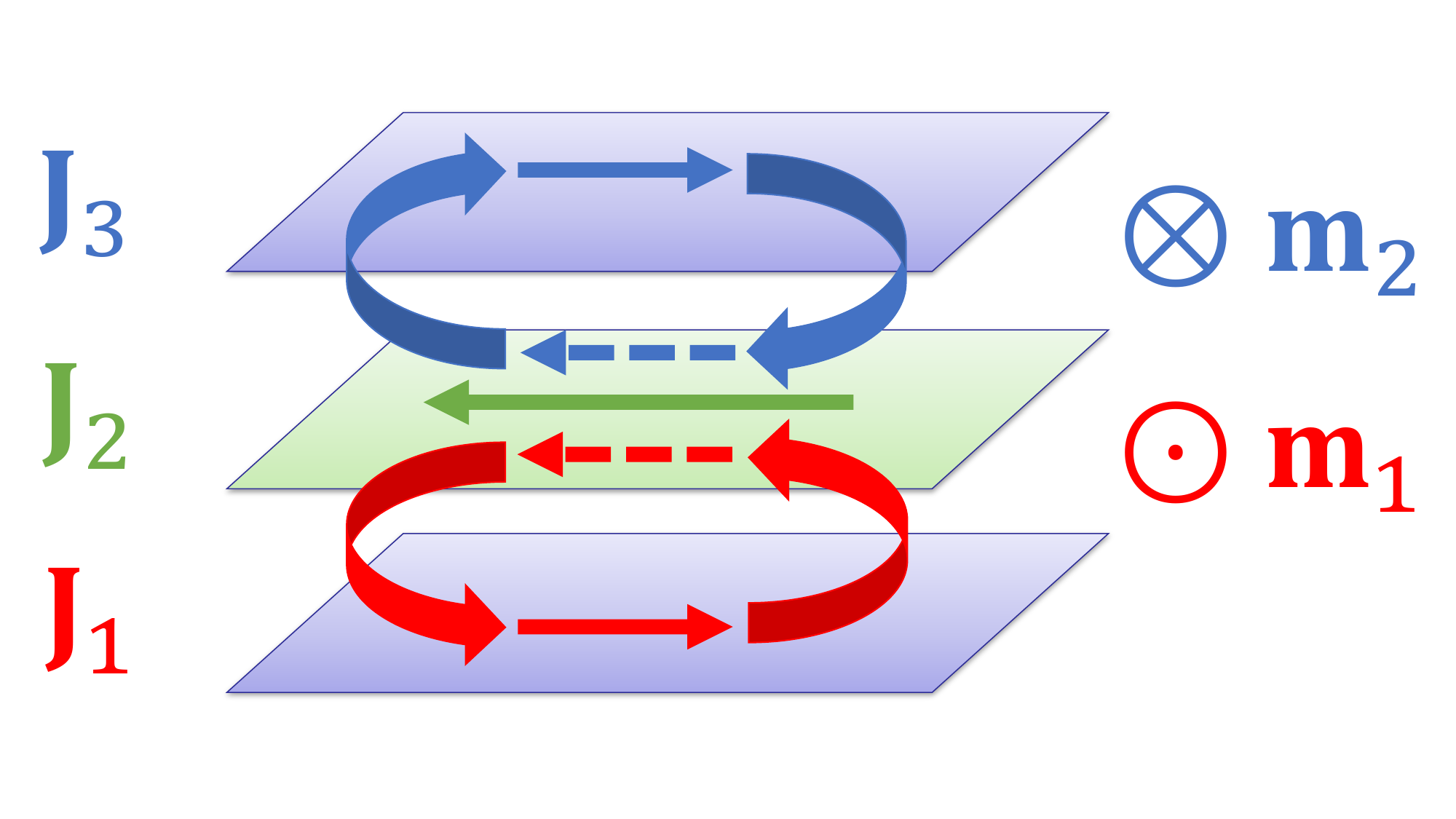}
\caption{Schematic view of the currents when only the even mode $\boldsymbol \cJ_+^-$ is present; see Eqs.~\eqref{eq:matrix-M} and~\eqref{eq:transformed-currents}. The solid arrows define the in-plane current densities $\J_1$, $\J_2$ and $\J_3$ with $\J_1=\J_3=-\J_2/2$. These currents give rise to the magnetic moments $\m_1$ and $\m_2$ with $\m_1=-\m_2$.}
\label{fig:Current}
\end{figure}

The  in-plane original current densities can always be described in terms of  electric and magnetic moments, as explained in Appendix~\ref{app:magnetization}. For the trilayer, the relations are
\begin{align} 
-i\omega\p&=\J_1+\J_2+\J_3\;,   \label{moments:1} \\
\frac{\m_2-\m_1}{a/6}&=\e_z\times(\boldsymbol \J_3+\boldsymbol \J_1-2 \boldsymbol\J_2)\;, \label{moments:2} \\
\frac{\m_1+\m_2}{a/2}&=\e_z\times(\boldsymbol \J_3 - \J_1)\;. \label{moments:3}
\end{align}
The linear combinations of currents in Eqs.~\eqref{moments:1}--\eqref{moments:3}  correspond to the modes $ \boldsymbol\cJ_+^+$, $\boldsymbol\cJ_+^- $, and $\boldsymbol\cJ_- $, respectively. 
In the absence of a net current, if $\p = 0 $, the in-plane current densities of modes $\{\boldsymbol\cJ_+^-, \boldsymbol\cJ_- \}$ can thus be thought of as magnetization currents associated with a uniform magnetization $\m_1$, filling the space between layers 1 and 2, and magnetization $\m_2$, filling the space between layers 2 and 3.
 Equations~\eqref{moments:2} and~\eqref{moments:3}  then follow, as shown in Appendix~\ref{app:magnetization}. 
 
For mirror-symmetric modes, when $\J_1=\J_3$, no net magnetic moment exists, i.e., $\m_1+\m_2=0$. 
Nevertheless, the finite value of $\m_1$ ($\m_2$) can be viewed as a gradient of the magnetic moment, as previously mentioned. The local magnetic moments of the mode $\boldsymbol \cJ_+^-$ are shown in Fig. \ref{fig:Current}. This mode is even in the current density, but odd in the magnetic moment. 

Therefore, by introducing the vertical gradients of the magnetic field and moment as
\begin{align}
\Delta\B_\parallel&=(\B_\parallel^{23}-\B_\parallel^{12})/2\;,\\
\Delta\m&=\m_2-\m_1\;,
\end{align}
and noticing that
\begin{align}
\B_\parallel&=(\B_\parallel^{23}+\B_\parallel^{12})/2\;,\\
\m_\parallel&=\m_1+\m_2\;,
\end{align}
we can formulate the following constitutive law:
\begin{widetext}
\begin{align}
\begin{pmatrix}
\p \\
\Delta\m	 \\  
\m_\parallel
\end{pmatrix}
=
\begin{pmatrix}
-3 \sigma_+^+/(i\omega) &a\left\{\sigma_+^{+-}(\e_z\times.)-\sigma_{xy}^{12}\right\}  &0\\
a\left\{\sigma_+^{+-}(\e_z\times.)+\sigma_{xy}^{12}\right\} &i\omega (a^2/6)\sigma_+^-	&0\\
0 &0 &i\omega (a^2/4) \sigma_{c}
\end{pmatrix}
\begin{pmatrix}
\E_\parallel \\
\Delta\B \\
\B_{\parallel}	
\end{pmatrix}\;.
\end{align}
\end{widetext}
Finally, we should point out that although the system is mirror symmetric and does not exhibit optical activity,  the $xy$-component, $\sigma_{xy}^{12}$, of the response introduces a local ``chiral coupling'' between the electric field and the gradient of the magnetic field.


\section{In-plane magnetic response}
\label{subsec:eff-parameters}
In this section, we derive a simplified formula for the
in-plane magnetic response, and estimate it as negligibly small. 
As shown in Sec.~\ref{subsec:in-plane-magn}, the in-plane magnetic response is given by
\begin{align}\label{eq:magnetic-resp}
\m_\parallel=i\omega\frac{a^2}{4}\sigma_{c}\B_\parallel\;.
\end{align}
We repeat that the magnetic-response parameter $\sigma_{c} $ cannot be read from the responses of the isolated   single-layer and twisted bilayer systems. However, mainly because of the kinematic constraints arising from the largely different Fermi velocities in these two systems, we expect $\sigma_{c} $  to be practically zero. In this section, we verify this property analytically.

The small, almost vanishing in-plane magnetic response of twisted trilayer graphene is in stark contrast to the large in-plane magnetic response of the twisted bilayer graphene~\cite{Stauber18}. The latter can even diverge and give rise to a Condon instability~\cite{Guerci21,Stauber23}. 

\subsection{Microscopic derivation and localization ansatz}
\label{subsec:micro-derivation}
We will derive an explicit, simplified formula for the in-plane magnetic response, 
Eq.~\eqref{eq:magnetic-resp}, near charge neutrality for alternating twists close to the magic angle. For this purpose, we introduce the following parameter of the in-plane magnetic response:
\begin{align}\label{eq:magnetic-resp-c}
\chi_{m}^c(\omega)=i\omega\frac{a^2}{4}\sigma_{c}(\omega)\;.
\end{align}
We need to estimate the coupling term $\sigma_{c}$, or response $\chi_m^c$. For this task, we employ a localization ansatz for the electronic Bloch wave functions of the effective twisted bilayer system at nearly flat bands, in the spirit of Ref.~\onlinecite{Sanchez24}.

Consider the eigenstates of Eq.~\eqref{eq:G-transf-vector}. For the effective bilayer system, our localization hypothesis reads~\cite{Sanchez24}
\begin{align}\label{eq:ansatz-TBG}
\begin{pmatrix}
	\langle \bx \vert b \rangle_1 \\
	\langle \bx \vert b\rangle_2 \\
\end{pmatrix}	
\simeq  e^{i(\K+\bq)\cdot x} \Phi_n(\sqrt{\xi}\bx/L_m)
\begin{pmatrix}
	P_{1,n} \\
	P_{2,n} \\
\end{pmatrix}
\end{align}
near the $K$ point. Here, $P_{\ell,n}$ is a $2\times 1$ vector for the sublattice polarization in each layer ($\ell=1,\,2$),  $n$ is the band index (e.g., $n=\pm 1$ for flat bands), $L_m$ is the moir\'e length ($|\K| L_m\gg 1$), $\bq$ is the quasi-momentum vector ($q\ll |\K|$), and $\xi\simeq 3$~\cite{Sanchez24}. The function $\Phi_n$ is localized in the moir\'e cell. By normalizing formula~\eqref{eq:ansatz-TBG} with $|P_{1,n}|\sim |P_{2,n}|$ we obtain
\begin{equation}
P_{\ell,n} \sim e^{i\vartheta_{\ell,n}}\sqrt{\frac{\xi}{4L_m^2}}\begin{pmatrix}
	1 \\
	1\\
\end{pmatrix}\;,	
\end{equation}
where $\vartheta_{\ell,n}$ is a phase that does not affect our results, and $\Phi_n$ is normalized to unity. The symbol $\sim$ means that numerical factors roughly equal to unity or smaller are ignored  for the sake of order-of-magnitude estimates. The eigenvectors of the single-layer system are 
\begin{align}\label{eq:states-SLG}
	\langle \bx \vert s\rangle =\frac{1}{\sqrt{2 A_m}} 
	e^{i(\K+\bq)\cdot x} 
\begin{pmatrix}
	1 \\
	\pm e^{i\theta/2}\frac{q_x+iq_y}{q} \\
\end{pmatrix}\;
\end{align}
for its two bands, where $A_m\sim L_m^2$.

Let us invoke Eqs.~\eqref{eq:ansatz-TBG} and~\eqref{eq:states-SLG} in Eq.~\eqref{eq:sigma-coupling}. We assume that $\omega < v_F q_m$ with $q_m=\frac{2\pi}{L_m}$. By only considering the nearly flat bands for the effective bilayer system, and relatively large  energies of the effective monolayer system so that $\omega_b-\omega_s\simeq -\omega_s$, we have 
\begin{align}
\chi_m^c\sim & \frac{g_v g_s}{2(4\pi)^2} v_F^2 a^2\frac{e^2}{\hbar}\sum_{n,n_s=\pm 1}|\bar C_n(\xi)|^2\int_{mBZ}d\bq 	\notag\\
&\times [f(E_n^b)-f(E_{n_s})]\,\frac{n_s v_F q}{(v_F q)^2-(\omega+i\delta)^2}\,.
\end{align}
In the above, the integration region, $mBZ$, is the moir\'e Brillouin zone, $E_n^b(q)$ ($E_{n_s}(q)$) is the energy of the effective bilayer (single-layer) system at band $n$ ($n_s$), and 
\begin{align*}
	\bar C_n(\xi)= \frac{1}{\sqrt{\xi}}\ \int d\bx' \,\Phi_n(\bx')\qquad (n=\pm 1)
\end{align*}
over the transformed moir\'e cell by $\bx\mapsto \bx'=\tfrac{\sqrt{\xi}}{L_m}\bx$. The scaling of this integral with $\sqrt{\xi}$ ensures that $|\bar C_n|\lesssim 1$ regardless of the value of $\xi>0$. 

\subsection{Estimate for the magnetic response}
\label{subsec:estimate_magnetic}
We will now evaluate the integral for the parameter $\chi_m^c$ by neglecting terms proportional to or smaller than $\tfrac{\omega}{v_F q_m}\ln(\frac{v_F q_m}{\omega})$; thus, $\hbar \omega$ may not exceed a few hundred meV which poses no practical restriction.

To demonstrate that $\chi_m^c$ is negligibly small, we compare it to $\chi_0=\hbar^{-2}e^2 t_G a_C^2$, which is used in discussions about lattice effects on the out-of-plane magnetic susceptibility in the single-layer graphene~\cite{Gomez-Santos2011}. Here, $a_C$ is the carbon-carbon distance and $t_G$ is the hopping energy for monolayer graphene. Hence, we compute
\begin{align}\label{eq:mag-resp-approx}
	\frac{\chi_m^c}{\chi_0}\sim \frac{3}{8}g_v g_s \langle f\rangle_m^c\,\frac{a^2/a_C}{L_m}\simeq 1.2\, \langle f\rangle_m^c\;
\end{align}
with $a= 1\,\rm{nm} \simeq 7a_C$, $L_m=a_C \sqrt{3}\sqrt{3i^2+3i+1}\simeq 61.43 a_C$ for $i=20$,  regarding the trilayer configuration (Fig.~\ref{fig:Geometry}). Here, we assume that the alternating-twist trilayer  is in the magic-angle regime.

Notice the linear scaling of $\tfrac{\chi_m^c}{\chi_0}$ with the inverse moir\'e length.  The positive factor $\langle f\rangle_m^c$ depends on the $|\bar C_n|$'s of nearly flat bands and the Fermi-Dirac distribution, $f$; and expresses a weighted radial-momentum average of $f(E_{-}(q))-f(E_+(q))$ for the single-layer graphene energies $E_{\pm}(q)=\pm \hbar v_F q$. This average is defined by
\begin{align*}
\langle f\rangle_m^c &= \tfrac{1}{2}(|\bar{C}_1|^2+|\bar{C}_{-1}|^2)\notag\\
&\times \left\{\frac{1}{q_m} \int_0^{q_m}dq\,[f(E_{-}(q))-f(E_+(q))]	\right\}\;.
\end{align*}
Note that $\langle f\rangle_m^c < 1$.

Equation~\eqref{eq:mag-resp-approx} suggests that the in-plane magnetic response of the alternating twisted trilayer is very weak, since it is comparable to or smaller than the corresponding atomistic (lattice) effect in monolayer graphene. Evidently, the response  function $\chi_m^c(\omega)$ has a plateau in frequency for  $\omega\lesssim v_F q_m$. The extension of our analysis to higher bands of the effective bilayer system might be pursued with a similar localization ansatz~\cite{StauberKohler2016}.

\section{Conclusion}
\label{sec:conclusion}
In this paper, we analytically described the overall optical response of the alternating-twist trilayer system. By performing a unitary transformation that decouples the trilayer Hamiltonian into an effective bilayer and a monolayer Hamiltonian, we showed that the optical response is basically given by the (known) responses of the bilayer and monolayer systems. However, the coupled response of the bilayer and single-layer current densities also appears, and this is proportional to the in-plane magnetic response. This response is usually very weak and we verified this property by an estimate in the magic-angle regime. Our analysis can explain the practical absence of an in-plane orbital magnetic susceptibility in the alternating-twist trilayer graphene~\cite{Park21,Zeyu21}. 

We further applied an alternative unitary transformation that involves electric and {\em local} magnetic moments. In this way, we formulated coupled constitutive equations based on the total electric field and the local magnetic fields. Nevertheless, this magneto-electric coupling does not lead to optical activity, as no real chirality can persist in this mirror-symmetric setting.

The results in this paper motivate further studies on the electromagnetic response of twisted multilayer systems. In particular, the response of tetralayer systems will be the subject of future studies.

\acknowledgments{The authors wish to thank E. Kaxiras, M. Luskin, J. T. Waters and Z. Zhu for useful discussions. The work of T.S. was supported by grant PID2020-113164GBI00 funded by MCIN/AEI/10.13039/501100011033, PID2023-146461NB-I00 funded by Ministerio de Ciencia, Innovaci\'on y  Universidades as well as by the CSIC Research Platform on Quantum Technologies PTI-001. G.G-S. acknowledges support from the Spanish Ministry of Science, Innovation and Universities  through the "María de Maeztu" Programme for Units of Excellence in R\&D (CEX2023-001316-M).}

  
  
 \appendix

\section{On time reversal symmetry}
\label{app:time-reverse}
In this appendix, we discuss implications of time reversal symmetry for the surface conductivity. The total electronic Hamiltonian, ${\mathcal H}$, of the multi-layered system in real space thus satisfies ${\mathcal H}={\mathcal H}^*$. 

We show that a time reversal ($\mathcal T$-) symmetric ${\mathcal H}$ implies a symmetric conductivity matrix, in the sense that interchanging the layer and polarization index pairs leaves the matrix invariant. In particular, we show that  (for $\nu, \nu'= x, y$ and $\ell, \ell'=1, 2, 3$) 
\begin{equation}\label{eq:cond-time-reversal}
	\sigma^{\ell\ell'}_{\nu\nu'}(\omega)=\sigma_{\nu'\nu}^{\ell'\ell}(\omega)~,\ \mbox{every}\ \omega\in \mathbb{C}.
\end{equation}

To prove Eq.~\eqref{eq:cond-time-reversal}, we start with the fact that there exists an anti-unitary transformation, ${\mathfrak U}$, such that for each eigenvector $\vert \psi\rangle$ of ${\mathcal H}$ we have the invariance property $\vert \psi \rangle = {\mathfrak U} \vert \psi \rangle$. We  apply ${\mathfrak U}^{-1} j_{0\nu}^\ell {\mathfrak U}=- j_{0\nu}^\ell$, where $j_{0\nu}^\ell$ denotes the $\nu$-directed  initial current of the $l$-th layer.

The trace of the Kubo formulation yields the following formula for the layer-resolved surface conductivity:
\begin{align*}
	\sigma_{\nu \nu'}^{\ell\ell'}(\omega)&=-\frac{e^2}{\hbar (\omega+i0^+)} \sum_{\{\vert\psi\rangle, \vert\psi'\rangle\}}\frac{f(E_\psi)-f(E_{\psi'})}{i (\omega+i0^++\omega_{\psi\psi'})}\notag\\
	& \qquad \times \langle \psi \vert {\mathfrak U}^{-1}j_{0,\nu}^\ell {\mathfrak U}\vert \psi'\rangle^* \, \langle \psi' \vert {\mathfrak U}^{-1}j_{0,\nu'}^{\ell'}{\mathfrak U}\vert \psi\rangle^*\notag\\
	&=-\frac{e^2}{\hbar (\omega+i0^+)} \sum_{\{\vert\psi\rangle, \vert\psi'\rangle\}}\frac{f(E_\psi)-f(E_{\psi'})}{i (\omega+i0^++\omega_{\psi\psi'})}\notag\\
	&\qquad \times \langle \psi \vert (j_{0,\nu'}^{\ell'})^\dagger\vert \psi'\rangle\,\langle \psi' \vert (j_{0,\nu}^\ell)^\dagger \vert \psi\rangle\quad (\mbox{real}\ \omega)\;. 
\end{align*}
Here, $E_\psi$ is the energy (eigenvalue of $\mathcal H$) corresponding to the normalized eigenvector $\vert \psi\rangle$ of $\mathcal H$, $f(E)$ is the Fermi-Dirac distribution, and $\omega_{\psi\psi'}=(E_\psi-E_{\psi'})/\hbar$. 
Relation~\eqref{eq:cond-time-reversal} follows by the Hermiticity of the current. This result can be analytically continued to complex $\omega$.

In our model, the Hamiltonian is given approximately near the  $K$- or $K'$-valley. Each of these Hamiltonians is not $\mathcal T$-symmetric. Since the total Hamiltonian is $\mathcal T$-symmetric, we apply the symmetrization
\begin{equation}\label{eq:cond-symmetrize}
\sigma_{\nu\nu'}^{\ell\ell'}(\omega)=(\sigma_{\nu\nu', K}^{\ell\ell'}(\omega)+\sigma_{\nu'\nu, K}^{\ell'\ell}(\omega))/2\;,	
\end{equation}
where $\sigma_{\nu\nu', K}^{\ell\ell'}$ is due to the $K$ valley. 
The resulting conductivity formulas will include the valley degeneracy factor, $g_v=2$; and a spin degeneracy factor, $g_s=2$.

 \section{Hamiltonian of trilayer system and unitary transformation to direct sum}
 \label{app:Ham-3layer} 
 
 In this appendix, we describe the Dirac Hamiltonian of the alternating-twist trilayer system in real space; and transform it unitarily into a direct sum of effective twisted bilayer and single-layer Hamiltonians, following Ref.~\onlinecite{Khalaf19}. Our setting lacks mirror symmetry. 

\subsection{Model Hamiltonian and moir\'e  potentials}
\label{subsec:Ham-3l}
The electron Schr\"odinger state vector is of the form
\begin{equation*}
	\psi= (\{\psi_\ell^s\})^T=(\psi_1^A\quad \psi_1^B\quad \psi_2^A\quad \psi_2^B\quad \psi_3^A\quad \psi_3^B)^T~.
\end{equation*}
 The twist angle of layer $\ell$ is $\theta_\ell=(-1)^\ell \theta/2$ ($\ell=1,\,2,\,3$). In the sublattice-layer representation, the unperturbed Hamiltonian of the $K$ valley reads
\begin{equation}\label{eq:H-trilayer}
\mathcal H^{tri}_K=
\begin{pmatrix}
{\mathfrak D}(\theta) & \mathfrak T^{12}(\bx) & 0\\
\mathfrak T^{12}(\bx)^\dag  & {\mathfrak D}(-\theta) & {\mathfrak T}^{23}(\bx) \\
0 & {\mathfrak T}^{23}(\bx)^\dag & {\mathfrak D}(\theta) \\
\end{pmatrix}\;;
\end{equation}
$\bx=(x,y)^T$ is the position vector in the reference plane ($\bx\in\mathbb{R}^2$).
The matrix-valued Dirac operator is 
\begin{equation}\label{eq:D-theta}
{\mathfrak D}(\theta)=v_F \hbar
\begin{pmatrix}
	0 & -2i e^{-i\frac{\theta}{2}}\px\\
-2i e^{i\frac{\theta}{2}}\bpx & 0 \\	
\end{pmatrix}\;,
\end{equation}
where $v_F\simeq 10^6$ m/s is the Fermi velocity of graphene, $\px=(\partial_x-i \partial_y)/2$ and $\bpx=(\partial_x+i\partial_y)/2$ in the $xy$ plane.

The related moir\'e potentials are described by
\begin{equation}\label{eq:m-moire-pot}
{\mathfrak T}^{\ell\ell'}(\bx)=
\begin{pmatrix}
	w_{AA}^{\ell\ell'} U_0^{\ell\ell'}(\bx) & w_{AB}^{\ell\ell'} U_1^{\ell\ell'}(\bx)\\
	w_{AB}^{\ell\ell'} {U_1^{\ell\ell'\ast}(-\bx)} & w_{AA}^{\ell\ell'} U_0^{\ell\ell'}(\bx)\\
\end{pmatrix}
\end{equation}
for $\ell'=\ell+1$ (if $\ell=1,\,2$), where $w_{AA}^{\ell\ell'}=\kappa w_{AB}^{\ell\ell'}$, $\kappa>0$, the coefficients $w_{AB}^{\ell\ell'}$ have units of energy and express interlayer tunneling, and 
\begin{equation}\label{eq:U-def}
U_\xi^{\ell\ell'}(\bx)=\sum_{n=1}^3 e^{-i \xi(n-1)\phi}\,e^{-i \boldsymbol q_n^{\ell\ell'}\cdot (\bx-\bD_{\ell\ell'})}
\end{equation}
with $\phi=2\pi/3$ and $\xi=0,\,1$. Here, we define~\cite{Khalaf19} 
\begin{align}
 \boldsymbol q_1^{\ell\ell'}&=2k_D\sin\biggl(\frac{\theta_{\ell'\ell}}{2}\biggr)\,\boldsymbol{\mathcal R}_{\phi_{\ell\ell'}}\cdot (0,-1)^T, \label{eq:q1-BLG}\\
 \boldsymbol q_{2,3}^{\ell\ell'}&=\boldsymbol{\mathcal R}_{\pm\phi} \boldsymbol q_1^{\ell\ell'}\;,\label{eq:q23-def}\\
 \bD_{\ell\ell'}&=\frac{\bd_\ell+\bd_{\ell'}}{2}+i \cot\biggl(\frac{\theta_{\ell'\ell}}{2} \biggr)	\,\boldsymbol\tau_y \cdot \frac{\bd_\ell-\bd_{\ell'}}{2}\;,
\label{eq:Dll}\\
 \theta_{\ell'\ell}&=\theta_{\ell'}-\theta_{\ell},\ \phi_{\ell\ell'}=\frac{\theta_{\ell}+\theta_{\ell'}}{2}\;, \label{eq:angles}\\
 \boldsymbol{\mathcal R}_{\phi}&=
 \begin{pmatrix}
\cos\phi & -\sin\phi \\
\sin\phi & \cos\phi\\	
\end{pmatrix} \notag\;,
\end{align}
where $k_D=|\K|=4\pi/(3\sqrt{3}a_C)$ is the Dirac momentum, $\boldsymbol \tau_y$ is the $y$-Pauli matrix, and $a_C\simeq 1.42$ \text{\AA} is the carbon-carbon distance.
The vector $\bd_\ell$ is the lateral shift of the $\ell$-th layer; and $\bD_{\ell\ell'}$ amounts to shifts in the potentials. We set $\bd_\ell=\bd=(d_1, d_2)^T$ for all $\ell$ by which $\bD_{\ell\ell'}=\bd$.

Next, we non-dimensionalize the model via 
the mapping $\boldsymbol x\mapsto \breve\bx=2k_D \sin(\theta/2) \bx$~\cite{Khalaf19}. Let the scaled position and shift vectors be $\breve \bx$ and $\breve{\!\boldsymbol d}$. The Hamiltonian $\mathcal H_K^{tri}$ exhibits the energy scale $2v_F\hbar k_D\sin(\theta/2)$; $\mathcal H_K^{tri}\mapsto \breve{\mathcal H}_K^{tri}=\big(2v_F\hbar k_D\sin(\theta/2)\big)^{-1}\mathcal H_K^{tri}$. We define
\begin{align*}
\alpha_{\ell\ell'}=\frac{w_{AB}^{\ell\ell'}}{2v_F \hbar k_D\sin(\theta/2)}>0\;,\quad \ell'=\ell+1\ (\ell=1,\,2)\;.\end{align*}
For ease of notation, we remove the breve symbol from $\breve \bx$, $\breve{\!\boldsymbol d}$ and $\breve{\mathcal H}_K^{tri}$, thus using $\bx$, $\bd=(d_1,d_2)^T$ and $\mathcal H_K^{tri}$ as the moir\'e-scaled variables; ditto for the potentials $\mathfrak T^{\ell\ell'}$.

 Hence, we now have the dimensionless potentials
\begin{align*}
{\mathfrak T}^{12}(\bx)&= 
\begin{pmatrix}
	\kappa \alpha_{12} U_0(\bx) & \alpha_{12} U_1^*(-\bx)\\
	\alpha_{12} U_1(\bx) & \kappa\alpha_{12} U_0(\bx) \\
\end{pmatrix},\notag\\
{\mathfrak T}^{23}(\bx)&\simeq  
\begin{pmatrix}
	\kappa \alpha_{23} U_0^*(\bx) & \alpha_{23} U_1^*(\bx)\\
	\alpha_{23} U_1(-\bx) & \kappa\alpha_{23} U_0^*(\bx) \\
\end{pmatrix}.
\end{align*}
If $\alpha_{12}\neq \alpha_{23}$ the ensuing Hamiltonian may lack mirror symmetry. This symmetry is recovered if $\alpha_{12}=\alpha_{23}$.

In view of $|\bd|\ll 1$, we have set 
\begin{align*}
	U_0(\bx)&=U_0^{12}(\bx)\;,\quad  U_0^{12\ast}(\bx)=U_0^*(\bx)\;,\notag\\
	U_0^{23}(\bx)&=U_0^*(\bx)\;,\quad U_0^{23\ast}(\bx)=U_0(\bx)\;,\notag\\
	U_1(\bx)&= U_1^{12\ast}(-\bx)\;,\quad U_1^{12}(\bx)=U_1^\ast(-\bx)\;,\notag\\
	U_1^{23}(\bx)&\simeq U_1^\ast(\bx)\;,\quad U_1^{23\ast}(-\bx)\simeq U_1(-\bx)\;.
\end{align*}
In the above, the first and fifth equations introduce $U_0$ and $U_1$, while the last two equations are implied by the condition $|\bd|\ll 1$.

For the sake of clarity, we give the explicit formulas for $U_0$ and $U_1$ at the moir\'e scale. These are
\begin{align}
U_0(\bx)&=e^{i (y-d_2)}+e^{-i \left(\frac{\sqrt{3}}{2}(x-d_1)+\frac{1}{2}(y-d_2)\right)}\notag\\
&\quad +e^{-i  \left(-\frac{\sqrt{3}}{2}(x-d_1)+\frac{1}{2}(y-d_2) \right)}	\notag\\
&\simeq e^{i y}+e^{-i \left(\frac{\sqrt{3}}{2}x+\frac{1}{2}y\right)}+e^{-i\left(-\frac{\sqrt{3}}{2}x+\frac{1}{2}y \right)}\;,\\
U_1(\bx)&\simeq e^{i y}+e^{i \frac{2\pi}{3}}e^{-i \left(\frac{\sqrt{3}}{2}x+\frac{1}{2}y\right)}\notag\\
&\quad +e^{-i \frac{2\pi}{3}}e^{-i \left(-\frac{\sqrt{3}}{2}x+\frac{1}{2}y \right)}\;.
\end{align}

The Hamiltonian ${\mathcal H}_{K'}^{tri}$ of the $K'$ valley comes from the complex conjugation of ${\mathcal H}^{tri}_K$ above. Thus, we write
\begin{equation*}
	{\mathcal H}_{K'}^{tri}=({{\mathcal H}_K^{tri}})^*\;,
\end{equation*}
in real space. Thus, the eigenvector (basis) sets of the two Hamiltonians are related by an anti-unitary transformation. The total contribution to the conductivity from both valleys under time reversal symmetry in linear response can be extracted from the $K$-valley Hamiltonian by a simple transposition. One must then symmetrize the optical conductivity (Appendix~\ref{app:time-reverse}).  

\subsection{Unitary transformations and direct sum}
\label{subsec:unitary}
Next, we outline the steps of a unitary transformation on $\mathcal H^{tri}_K$, which eventually yields the mapping
\begin{equation}\label{eq:mapp-dir-sum}
	{\mathcal  H}^{tri}_K \,\mapsto \,{\mathcal H}^{eff}_{K}={\mathfrak S}^\dag\, {\mathcal H}^{tri}_K \,{\mathfrak S}={\mathcal H}_{\lambda,K}^{(2)} \oplus {\mathcal H}^{(1)}_K~, 
\end{equation}
where ${\mathcal H}_{\lambda,K}^{(2)}$ is the effective, $K$-valley twisted bilayer Hamiltonian with parameter $\lambda=\sqrt{\alpha_{12}^2+\alpha_{23}^2}$, and ${\mathcal H}^{(1)}_K$ is the respective single-layer Hamiltonian, as explained below. Hence, we can write any eigenvector of ${\mathcal H}^{eff}_{K}$ in the form $(\vert b\rangle_1, \vert b\rangle_2, 0)^T$ or $(0,0, \vert s\rangle)^T$ where $(\vert b\rangle_1, \vert b\rangle_2)^T$ and $\vert s\rangle$ are eigenvectors of ${\mathcal H}_{\lambda,K}^{(2)}$ and 
${\mathcal H}^{(1)}_K$, respectively. We have
\begin{equation*}
	{\mathcal H}_{\lambda,K}^{(2)}\begin{pmatrix}
\vert b\rangle_1 \\
\vert b\rangle_2	
\end{pmatrix}
=E_{b}^{(2)} \begin{pmatrix}
\vert b\rangle_1 \\
\vert b\rangle_2	
\end{pmatrix}\;,\quad  {\mathcal H}^{(1)}_K\vert s\rangle = E_s^{(1)}\vert s \rangle\;;
\end{equation*}
$E_{b,s}^{(2,1)}$ is the eigenvalue of the Dirac Hamiltonian for the twisted bilayer ($b$) and single-layer ($s$) system. Each of the symbols $s$ and $b$ stands for the combined band index and continuum quasi-momentum variable of the scaled Brillouin zone. We set $E_s=E_s^{(1)}$ and $E_b=E_b^{(2)}$.

We will show that in the sublattice-layer form~\cite{Khalaf19}
\begin{equation}\label{eq:T-matrix}
{\mathfrak S}=\begin{pmatrix}
\displaystyle \alpha_{12}/\lambda & 0 & \displaystyle \alpha_{23}/\lambda \\
0 & 1 & 0 \\
\displaystyle \alpha_{23}/\lambda & 0 & \displaystyle -\alpha_{12}/\lambda\\	
\end{pmatrix}\;.
\end{equation}
Note that $\mathfrak S$ is represented by a $6\times 6$ matrix. In the case with mirror symmetry, if $\alpha_{12}=\alpha_{23}$, we have 
\begin{equation}\label{eq:T-matrix-mirror}
{\mathfrak S}=\tfrac{1}{\sqrt{2}}\begin{pmatrix}
1 & 0 & 1 \\
0 & \sqrt{2} & 0 \\
1 & 0 &  -1\\	\end{pmatrix}\;.
\end{equation}

This unitary transformation is described as a result of four basic, successive unitary operations, as detailed below~\cite{Khalaf19}. 
To simplify notation, we will omit the valley index ($K$),  unless we state otherwise. 

\subsubsection{Layer ordering}
\label{sssec:layer-tr}
We apply ${\mathcal H}^{tri}\mapsto \overset{\,\circ}{{\mathcal H}}:=\Omega^\dag {\mathcal H}^{tri}\Omega $ with state vector
\begin{align*}
	 \psi=& (\psi_1^A\quad \psi_1^B\quad \psi_2^A\quad \psi_2^B\quad \psi_3^A\quad \psi_3^B)^T\notag\\
	&\mapsto \overset{\,\circ}{\psi}=\Omega^\dag \psi = (\psi_1^A\quad \psi_2^A\quad \psi_3^A\quad \psi_1^B\quad \psi_2^B\quad \psi_3^B)^T\;.
\end{align*}
The transformation operator $\Omega$ in matrix form is
\begin{equation*}
	\Omega=\begin{pmatrix}
1 & 0 & 0 & 0 & 0 & 0 \\
0 & 0 & 0 & 1 & 0 & 0 \\	
0 & 1 & 0 & 0 & 0 & 0 \\
0 & 0 & 0 & 0 & 1 & 0 \\
0 & 0 & 1 & 0 & 0 & 0\\
0 & 0 & 0 & 0 & 0 & 1\\
\end{pmatrix}~.
\end{equation*}

\subsubsection{Layer alignment }
\label{sssec:layer-tr}
We now apply $\overset{\,\circ}{{\mathcal H}}\mapsto \check{{\mathcal H}}=Y^\dag \overset{\circ}{{\mathcal H}} Y$ so that
\begin{align*}
\overset{\,\,\circ}{\psi}=&(\psi_1^A\quad \psi_2^A\quad \psi_3^A\quad \psi_1^B\quad \psi_2^B\quad \psi_3^B)^T \notag\\
&\mapsto \check {\psi}=Y^\dag \overset{\,\,\circ}{\psi}=(\psi_1^A\quad \psi_3^A\quad \psi_2^A\quad \psi_1^B\quad \psi_3^B\quad \psi_2^B)^T\;;
\end{align*}
\begin{equation*}
	Y=\begin{pmatrix}
	1 & 0 & 0 & 0 & 0 & 0\\
	0 & 0 & 1 & 0 & 0 & 0 \\
	0 & 1 & 0 & 0 & 0 & 0 \\
	0 & 0 & 0 & 1 & 0 & 0 \\
	0 & 0 & 0 & 0 & 0 & 1 \\
	0 & 0 & 0 & 0 & 1 & 0\\
\end{pmatrix}\;.
\end{equation*}
Thus, we have~\cite{Khalaf19}
\begin{equation*}
 \check{{\mathcal H}}=
 \begin{pmatrix}
	{\mathcal M} & {\mathcal D}^\dag \\
	{\mathcal D} & {\mathcal M}\\
	\end{pmatrix}\;,
\end{equation*}
\begin{align*}
	{\mathcal M}&=\kappa 
	\begin{pmatrix}
		0 & 0 & \alpha_{12} U_0(\bx) \\
		0 & 0 & \alpha_{23} U_0(\bx) \\
		\alpha_{12} U_0^*(\bx) & \alpha_{23} U_0^*(\bx) & 0 \\ 
	\end{pmatrix}\notag\\
	&=\kappa 
	\begin{pmatrix}
	\bf{0} & {W} U_0(\bx) \\
	{W}^T U_0^*(\bx) & 0 \\	
	\end{pmatrix}\;,\quad   
	{W}=
	\begin{pmatrix}
		\alpha_{12}\\
		\alpha_{23}
	\end{pmatrix}\;.
\end{align*}
Here, $\mathcal D$ is represented by the following $3\times 3$ matrix:
\begin{equation*}
	{\mathcal D}=
	\begin{pmatrix}
		-2i e^{i\theta/2} \bpx  & W U_1(\bx) \\
		W^T U_1(-\bx) & -2i e^{-i\theta/2}\bpx  \\
	\end{pmatrix}\;.
\end{equation*}

\subsubsection{Singular-value decomposition}
\label{sssec:sing-val}
Next, we apply $\check{{\mathcal H}}\mapsto \wideparen {{\mathcal H}}=V^\dag \check{{\mathcal H}} V$ where 
\begin{equation*}
	V=
	\begin{pmatrix}
		{\mathcal V} & 0\\
		0 & {\mathcal V}\\
	\end{pmatrix}\;,\quad {\mathcal V}=\mathrm{diag}({\mathcal A}, \mathcal B)\;,
\end{equation*}
and the $2\times 2$ matrix ${\mathcal A}$ and scalar $\mathcal B$ stem from 
$W={\mathcal A} \Lambda \mathcal B^*$
with $\Lambda=(\lambda, 0)^T$ and  $\lambda=\sqrt{W^T W}=\sqrt{\alpha_{12}^2+\alpha_{23}^2}$.
By a direct computation, we have 
\begin{equation*}
	{\mathcal A}=
	\begin{pmatrix}
		\alpha_{12}/\lambda & \alpha_{23}/\lambda \\
		\alpha_{23}/\lambda & -\alpha_{12}/\lambda \\
	\end{pmatrix}\;,\quad 
	\mathcal B=1\;.
\end{equation*}

The transformed Hamiltonian is thus written as 
\begin{widetext}
\begin{equation*}
	\wideparen{{\mathcal H}}=
	\begin{pmatrix}
		0 & 0 & \kappa\lambda U_0(\bx) & -2i e^{-i\theta/2}\partial & 0 & \lambda U_1^*(-\bx)\\
		0 & 0 & 0 & 0 & -2i e^{-i\theta/2} \partial & 0 \\
		\kappa\lambda U_0^*(\bx) & 0 & 0 & \lambda U_1^*(\bx) & 0 & -2i e^{i\theta/2}\partial \\
		-2i e^{i\theta/2}\bpx & 0 & \lambda U_1(\bx) & 0 & 0 & \kappa\lambda U_0(\bx)\\
		0 & -2i e^{i\theta/2}\bpx & 0 & 0 & 0 & 0 \\
		\lambda U_1(-\bx) & 0 & -2i e^{-i\theta/2}\bpx & \kappa\lambda U_0^*(\bx) & 0 & 0\\
	\end{pmatrix}\;.
\end{equation*}
\end{widetext}

\subsubsection{Conversion to direct sum}
\label{sssec:sing-val}
We apply 
$\wideparen{{\mathcal H}}\mapsto {\mathcal H}^{eff}=Z^\dag \wideparen{{\mathcal H}} Z={\mathcal H}_\lambda^{(2)}\oplus {\mathcal H}^{(1)}$ 
with 
\begin{equation*}
	Z=
	\begin{pmatrix}
		1 & 0 & 0 & 0 & 0 & 0\\
		0 & 0 & 0 & 0 & 1 & 0 \\
		0 & 0 & 1 & 0 & 0 & 0 \\ 
		0 & 1 & 0 & 0 & 0 & 0 \\
		0 & 0 & 0 & 0 & 0 & 1\\
		0 & 0 & 0 & 1 & 0 & 0 \\
	\end{pmatrix}.
\end{equation*}
In the above, ${\mathcal H}^{(1)}$ and ${\mathcal H}_\lambda^{(2)}$ denote the moir\'e-scaled effective single-layer and twisted bilayer Hamiltonians, respectively.  To simplify notation, set ${\mathcal H}^{(1)}={\mathcal H}_1$ and ${\mathcal H}_\lambda^{(2)}={\mathcal H}_2$. These operators read 
\begin{align}
	\mathcal H_1&=\mathfrak D(\theta)=
	\begin{pmatrix}
		0 & -2i e^{-i\theta/2}\partial \\
		-2i e^{i\theta/2}\bpx & 0\\
	\end{pmatrix} \label{eq:eff-H1-def}
\end{align}
and
\begin{align}
	\mathcal H_2&=
	\begin{pmatrix}
		\mathfrak D(\theta)  & \lambda \,\mathcal U(\bx) \\
		\lambda \,\mathcal U(\bx)^\dagger & \mathfrak D(-\theta) \\
	\end{pmatrix}\;,\label{eq:eff-H2-def}
\end{align}
where 
\begin{align*}
	\mathcal U(\bx)=
	\begin{pmatrix}
		\kappa U_0(\bx) & U_1^*(-\bx) \\
		U_1(\bx) & \kappa U_0(\bx) 
	\end{pmatrix}\;.
\end{align*}

In summary, the operator ${\mathfrak S}$ of Eq.~\eqref{eq:T-matrix}, which converts the $K$-valley  Hamiltonian ${\mathcal H}^{tri}_K$ to ${\mathcal H}^{eff}_K$, is computed directly by ${\mathfrak S}={\Omega} {Y} {V} {Z}$. 
The corresponding matrix is real. Hence, the same matrix is used to convert the $K'$-valley Hamiltonian, ${\mathcal H}^{tri}_{K'}=({\mathcal H}^{tri}_K)^*$ in real space, into a direct sum of effective twisted bilayer and single-layer Hamiltonians. 

 \section{Optical conductivity of alternating-twist trilayer system: Calculations}
 \label{app:conductivity}
In this appendix, we determine the optical conductivity for the alternating twisted trilayer system by the Kubo formulation without  mirror symmetry. The main assumptions are time reversal symmetry and spatial isotropy. We invoke the unitary transformation of the $K$-valley trilayer Hamiltonian into a direct sum of the effective Hamiltonians ${\mathcal H}_1$ and ${\mathcal H}_2$ (Appendix~\ref{app:Ham-3layer}). We use moir\'e-scaled quantities, so that the units of length and energy are $\{2k_D\sin(\theta/2)\}^{-1}$ and $2v_F\hbar k_D \sin(\theta/2)$. Thus, the (dimensionless) current operator in layer $\ell$ is $\boldsymbol j^\ell=-i[\bx^\ell, \mathcal H^{tri}]$, where $\mathcal H^{tri}$ and $\bx^\ell$ are the moir\'e-scaled trilayer Hamiltonian and vector position. We often suppress the subscript $K$ for the valley. 

\subsection{Methodology and general results}
\label{subsec:main-res-trilayer}
We transform the unperturbed trilayer Hamiltonian, $\mathcal H^{tri}_K$, by $\mathfrak S$  into the direct sum $\mathcal H_2\oplus \mathcal H_1$. Hence, we compute the trace of the current-current commutators on the basis formed by the eigenvectors of $\mathcal H_2\oplus \mathcal H_1$.

Consider the set of all state vectors of the form
\begin{align*}  
	\vert \underline{\psi}\rangle=\mathfrak G^\dagger  \vert\psi\rangle=\begin{pmatrix}
	\vert b \rangle_1 \\
	\vert b \rangle_2 \\
	 0  \\
	 \end{pmatrix}\;\ \mbox{or}\  	 
	 \begin{pmatrix}
	0 \\
	0 \\
	 \vert s \rangle \\
	 \end{pmatrix}\;,
	\end{align*}
which are eigenvectors of $\mathcal H_2\oplus \mathcal H_1$; 
$\vert b\rangle=(\vert b\rangle_1, \vert b\rangle_2)^T$ and $\vert s\rangle$ denote the effective twisted bilayer and single-layer Hamiltonian eigenvectors, with energies $E_b=E_b^{(2)}$ and $E_s=E_s^{(1)}$. The symbols $b$ and $s$ express the combined band index and quasi-momentum in the scaled moir\'e Brillouin zone.  

For real $\omega$, the $K$-valley Kubo formula yields
\begin{align}\label{eq:Kubo-1part-freq}
	&\sigma_{\nu\nu',K}^{\ell\ell'}(\omega)=C_0\frac{e^2}{\hbar(\omega+i\delta)}\sum_{\{\vert \underline{\psi}\rangle\}}f(E_\psi)\int_0^{\infty}dt\,e^{i(\omega+i\delta) t}\notag\\
	&\qquad\times \langle \underline{\psi} \vert [{\underline j}_{\nu,K}^{\ell}(t), \underline{j}_{\nu',K}^{\ell'}(0)]\vert \underline{\psi}\rangle\notag\\
	&=C_0\frac{e^2}{\hbar (\omega+i\delta)} \left\{\sum_{b}f_b \int_0^\infty dt\,e^{i(\omega+i\delta)t}\right.\notag\\
	&\quad\times ({}_1\!\langle b \vert, {}_2\!\langle b\vert) \left(V^{\ell\ell'}_{bb,\nu\nu'}(t)-   V^{\ell\ell'}_{bb,\nu\nu'}(t)^\dagger\right)
	\begin{pmatrix}
		\vert b\rangle_1 \\
		 \vert b\rangle_2 
	\end{pmatrix}\notag\\
	&+\sum_s f_s \int_0^\infty dt\,e^{i(\omega+i\delta)t}\notag\\
	&\quad\times \left.\langle s \vert V^{\ell\ell'}_{ss,\nu\nu'}(t)- V^{\ell\ell'}_{ss,\nu\nu'}(t)^\dagger\vert s\rangle \right\}\;,\quad \delta\downarrow 0\;,
	\end{align}
for $\ell,\, \ell'=1,\, 2,\, 3$ and $\nu,\, \nu'= x,\, y$. The $\nu$-directed current 
$\underline{j}_{\nu,K}^\ell(t)=\mathfrak{S}^\dagger j_{\nu,K}^\ell(t) \mathfrak{S}$, frequency $\omega$ and time $t$ are moir\'e scaled and dimensionless; and $C_0=\{4k_D^2\sin^2(\theta/2)A_m\}^{-1}=\tfrac{3\sqrt{3}}{8\pi^2}$ where $A_m=\tfrac{3\sqrt{3}}{8}\left(\tfrac{a_C}{\sin(\theta/2)}\right)^{2}$ is the area of the moir\'e cell and $a_C$ is the carbon-carbon distance. We set $f_{\eta}=f(E_{\eta})$ as the Fermi-Dirac distribution, and $V_{\eta\eta',\nu\nu'}^{\ell\ell'}(t)=\mathbf{e}_\nu \cdot \tvpb^{\ell\ell'}_{\eta\eta'}(t)\cdot\mathbf{e}_{\nu'}$ ($\e_{\nu}$: Cartesian unit vector) for $\eta,\,\eta'=b,\,s$. The matrix-valued operator $\tvpb^{\ell\ell'}_{\eta\eta'}(t)$ is defined by 
\begin{align}\label{eq:V-tensor-def}
	\underline{\boldsymbol j}^{\ell}_K(t) \underline{\boldsymbol j}_{0,K}^{\ell'}&=\begin{pmatrix}
	\tvpb_{bb}^{\ell\ell'}(t) & \tvpb_{bs}^{\ell\ell'}(t) \\
	\tvpb_{sb}^{\ell\ell'}(t) & \tvpb_{ss}^{\ell\ell'}(t)\\
\end{pmatrix}\;,
\end{align}
and thus vanishes for $t<0$; $\boldsymbol{j}_0^\ell=\boldsymbol {j}^\ell(0)$.
By Eq.~\eqref{eq:Kubo-1part-freq}, we only need the diagonal terms, $\tvpb_{\eta\eta}^{\ell\ell'}(t)$.
The symbol $\sum_{\eta}$ ($\eta=b, s$) means summation over band indices and integration in the quasi-momentum over the scaled Brillouin zone. The elements $\langle \eta \vert \cdot \vert \eta \rangle$ ($\eta=b, s$) of Eq.~\eqref{eq:Kubo-1part-freq} account for $\bx$-inner products of scalar Bloch wave functions, which compose $\langle \bx\vert\eta\rangle$.

In Eq.~\eqref{eq:Kubo-1part-freq}, each sum $\sum_\eta$ is replaced by $\sum_{\eta,\eta'}$ or $\sum_{\eta,\bar \eta'}$ by use of the resolution of the identity, $1=\sum_{\eta}\vert \eta \rangle \langle \eta\vert$, in the respective Hilbert space (for $\eta=b,\,s$). Here, we set $(\eta,\eta')=(b,b')$ or $(s, s')$, while $\bar \eta'$ in $(\eta,\bar\eta')$ takes values complementary to those of $\eta$ in the sense that $\bar\eta'=s'$ if $\eta=b$ and $\bar\eta'=b'$ if $\eta=s$. 

For every $\eta$ in these sums, the Fourier transform of $\langle\eta\vert V_{\eta\eta,\nu\nu'}^{\ell\ell'}(t)\vert \eta\rangle $ contributes the factors 
\begin{equation*}
	-\frac{1}{i (\omega+\omega_{\eta\eta'}+i\delta)}\quad \mbox{or}\quad -\frac{1}{i (\omega+\omega_{\eta \bar\eta'}+i\delta)}\;,
\end{equation*}
as $\delta\downarrow 0$.  The Fourier transform of $\langle\eta\vert V_{\eta\eta,\nu\nu'}^{\ell\ell'}(t)^\dagger\vert \eta\rangle $ contributes the respective factors 
\begin{equation*}
	-\frac{1}{i (\omega-\omega_{\eta\eta'}+i\delta)}\quad \mbox{or}\quad -\frac{1}{i (\omega-\omega_{\eta \bar\eta'}+i\delta)}\;.
\end{equation*}

The ensuing optical conductivity tensor, $\bsigma(\omega)$, comes from the symmetrization of the $K$-valley contribution (Appendix~\ref{app:time-reverse}).   This tensor $\bsigma$ reads
\begin{equation*}
	\bsigma(\omega)=\begin{pmatrix}
\bsigma^{11}(\omega) & \bsigma^{12}(\omega) & \bsigma^{13}(\omega) \\
\bsigma^{21}(\omega) & \bsigma^{22}(\omega) & \bsigma^{23}(\omega)\\
\bsigma^{31}(\omega) & \bsigma^{32}(\omega) & \bsigma^{33}(\omega)\\	
\end{pmatrix}\;,
\end{equation*}
where $\bsigma^{\ell\ell'}$ are $2\times 2$ matrices for the layer-resolved conductivities. We will express $\bsigma^{\ell\ell'}$ in terms of principal conductivities associated with two effective, simpler systems, for {\em arbitrary} positive parameters $\alpha_{12(23)}$ of the  trilayer Hamiltonian $\mathcal H^{tri}_K$ (Appendix~\ref{app:Ham-3layer}).  

After some algebra, we obtain  (see Sec.~\ref{subsec:cond-trilayer-details}) 
\begin{align}
	\bsigma^{11}&=\lambda^{-4}(\alpha_{12}^4\bsigma_{BL}^{11}+\alpha_{23}^4	\bsigma_{SL}+\alpha_{12}^2\alpha_{23}^2\bsigma_c)\label{eq:sigma11-tri}\\
&=\begin{pmatrix}
\sigma_0^{11} & 0\\
0 & \sigma_0^{11}\\	
\end{pmatrix}\;,\label{eq:sigma11-tri-diag-form}
\end{align}
\begin{align}\label{eq:sigma12-21-tri}
	\bsigma^{12}&=\frac{\alpha_{12}^2}{\lambda^2}\bsigma^{12}_{BL}\notag\\
	&=\begin{pmatrix}
\sigma_0^{12} & \sigma_{xy}^{12}\\
-\sigma_{xy}^{12} & \sigma_0^{12}\\	
\end{pmatrix}=(\bsigma^{21})^T\;,
\end{align}
\begin{align}
	\bsigma^{13}&=\frac{\alpha_{12}^2\alpha_{23}^2}{\lambda^4}
	(\bsigma_{BL}^{11}+\bsigma_{SL}-\bsigma_c)\label{eq:sigma13-31-tri} \\
&=\begin{pmatrix}
\sigma_0^{13} & 0\\
0 & \sigma_0^{13}\\	
\end{pmatrix}=\bsigma^{31}\;, \label{eq:sigma13-31-tri-diag-form}
\end{align}
\begin{equation}\label{eq:sigma22-tri}
	\bsigma^{22}=\bsigma_{BL}^{22}=
	\begin{pmatrix}
		\sigma_0^{22} & 0\\
0 & \sigma_0^{22}\\
	\end{pmatrix}\;,
\end{equation}
\begin{align}\label{eq:sigma23-32-tri}
	\bsigma^{23}&=\frac{\alpha_{23}^2}{\lambda^2}\bsigma^{21}_{BL}\notag\\
	&=\frac{\alpha_{23}^2}{\alpha_{12}^2}\bsigma^{21}=(\bsigma^{32})^T\;,
\end{align}
\begin{align}\label{eq:sigma33-tri}
	\bsigma^{33}&=\lambda^{-4}(\alpha_{23}^4\bsigma_{BL}^{11}+\alpha_{12}^4	\bsigma_{SL}+\alpha_{12}^2\alpha_{23}^2\bsigma_c)\notag\\
	&=\bsigma^{11}|_{\alpha_{12}\leftrightarrow \alpha_{23}}=\begin{pmatrix}
\sigma_0^{33} & 0\\
0 & \sigma_0^{33}\\	
\end{pmatrix}\;. 
\end{align} 
The first equation for $\bsigma^{33}$ comes from interchanging the constants $\alpha_{12}$ and $\alpha_{23}$ in $\bsigma^{11}$. In Eqs.~\eqref{eq:sigma11-tri}--\eqref{eq:sigma33-tri}, we express the layer-resolved conductivities in terms of the principal $2\times 2$ matrices $\bsigma_{BL}^{\varsigma\varsigma'}$, $\bsigma_{SL}$, and $\bsigma_c$ ($\varsigma,\,\varsigma'=1,\,2$).
Here, $\bsigma_{BL}^{\varsigma\varsigma'}$  and $\bsigma_{SL}$ are conductivities that separately arise from the effective twisted bilayer and single-layer Hamiltonians, respectively; cf. Eqs.~\eqref{eq:eff-H2-def} and~\eqref{eq:eff-H1-def} in Appendix~\ref{app:Ham-3layer}. On the other hand, $\bsigma_c$ couples the two effective systems. 

The principal matrices have the forms
\begin{align*}
\bsigma_{SL}&=\begin{pmatrix}
\sigma_e^{(1)} & 0 \\
0 & \sigma_e^{(1)}\\	
\end{pmatrix}\;,\\
	\bsigma_{BL}^{\varsigma\varsigma}&=\begin{pmatrix}
\sigma_{e}^{(2)} & 0 \\
0 & \sigma_{e}^{(2)}\\	
\end{pmatrix}\quad (\varsigma=1, 2)\;,\notag\\
\bsigma_{BL}^{12}&=\begin{pmatrix}
\sigma_e^{12} & \sigma_{e,xy}^{12}\\
-\sigma_{e,xy}^{12} & \sigma_e^{12}\\	
\end{pmatrix}=(\bsigma_{BL}^{21})^T\;,\\
\bsigma_c&=\begin{pmatrix}
\sigma_{c} & 0 \\
0 & \sigma_{c} \\	
\end{pmatrix}\;.
\end{align*}
The scalars $\sigma_e^{(1,2)}$, $\sigma_{e}^{12}$, $\sigma_{e,xy}^{12}$, and $\sigma_{c}$ respectively denote the in-plane conductivities of the effective single-layer and twisted bilayer systems, the covalent drag and chiral (or Hall) conductivities of the effective bilayer system,   and the coupling of the two  systems. Note that $\sigma_0^{22}=\sigma_e^{(2)}$ in Eq.~\eqref{eq:sigma22-tri}. In Sec.~\ref{subsec:cond-trilayer-details}, we express all these scalar functions of $\omega$ in terms of inner products of suitable pseudo-spin wave functions.

Hence, the elements $\sigma_0^{\ell\ell'}$ of Eqs.~\eqref{eq:sigma11-tri}--\eqref{eq:sigma33-tri} are  
\begin{align}
\sigma_0^{11}&=\lambda^{-4}(\alpha_{12}^4\sigma_{e}^{(2)}+\alpha_{23}^4	\sigma_{e}^{(1)}+\alpha_{12}^2\alpha_{23}^2\sigma_{c})\;,\label{eq:sigma0+}\\
\sigma_0^{12}&=\frac{\alpha_{12}^2}{\lambda^2}\sigma_{e}^{12}\;,\quad \sigma_{xy}^{12}=\frac{\alpha_{12}^2}{\lambda^2}\sigma_{e,xy}^{12}\;,\label{eq:sigma12-wo-mirror} \\
\sigma_0^{13}&=\frac{\alpha_{12}^2\alpha_{23}^2}{\lambda^{4}}(\sigma_{e}^{(2)}+\sigma_{e}^{(1)}-\sigma_{c})\;,\label{eq:sigma0-}\\
\sigma_0^{22}&=\sigma_e^{(2)}\;, \label{eq:sigma022-wo-mirror}\\
\sigma_0^{33}&=\lambda^{-4}(\alpha_{23}^4\sigma_{e}^{(2)}+\alpha_{12}^4	\sigma_{e}^{(1)}+\alpha_{23}^2\alpha_{12}^2\sigma_{c})\notag\\
&=\sigma_0^{11}|_{\alpha_{12}\leftrightarrow \alpha_{23}}\;. \label{eq:brevesigma0+}
\end{align}

In particular, let us consider $\alpha_{12}=\alpha_{23}$, by which mirror symmetry is recovered. We find 
\begin{align}\label{eq:sigma11-m}
	\bsigma^{11}&=\tfrac{1}{4}(\bsigma_{BL}^{11}+\bsigma_{SL}+\bsigma_c)\;,
\end{align}
\begin{align}\label{eq:sigma12-m}
	\bsigma^{12}&=\tfrac{1}{2}\bsigma^{12}_{BL}=(\bsigma^{21})^T\;,
\end{align}
\begin{align}\label{eq:sigma13-31-m}
	\bsigma^{13}&=\tfrac{1}{4}
	(\bsigma_{BL}^{11}+\bsigma_{SL}-\bsigma_c)=\bsigma^{31}\;,
\end{align}
\begin{align}\label{eq:sigma23-m}
	\bsigma^{23}&=\tfrac{1}{2}\bsigma^{21}_{BL}=\bsigma^{21}=(\bsigma^{32})^T\;,
\end{align}
\begin{equation}\label{eq:sigma33-m}
	\bsigma^{33}=\bsigma^{11}\;. 
\end{equation} 

\subsection{Trace algebra}
\label{subsec:vel-commutator-3l}
Next, we provide details on the layer-resolved conductivities. We compute the unitarily transformed currents and $\boldsymbol V_{\eta\eta'}^{\ell\ell'}(t)$ in view of Eq.~\eqref{eq:mapp-dir-sum} of Appendix~\ref{app:Ham-3layer} and time evolution under Hamiltonian $\mathcal H^{eff}_{K}$. 

The trace of the Kubo formula is computed on the basis formed by the eigenvectors of $\mathcal H^{eff}=\mathcal H_{2}\oplus \mathcal H_{1}$. The unitarily transformed currents are 
\begin{equation*}
\underline{j}^{\ell}_{0\nu}=\mathfrak S^\dag j^{\ell}_{0\nu} \mathfrak S~,\quad  \underline{j}^{\ell}_\nu(t)=e^{i \mathcal H^{eff} t} \underline{j}^{\ell}_{0\nu} e^{-i \mathcal H^{eff} t}\;.	
\end{equation*}
The initial original (non-transformed) $\nu$-directed current in layer $\ell$ is $j_{0\nu}^{\ell}=\boldsymbol j_0^{\ell}\cdot \e_\nu$ with $\boldsymbol j^{\ell}_{0}=-i[\bx^{\ell}, \mathcal H^{tri}]$, where $\e_{\nu}$ is a Cartesian unit vector in the $xy$-plane.  

Therefore, we start with the matrix representations
\begin{align*}
\boldsymbol j_0^{1}&=
\begin{pmatrix}
	\taub^{-\theta} & 0 & 0 \\
	0 & 0 & 0 \\
	0 & 0 & 0 \\
\end{pmatrix}\;,\quad 
\boldsymbol j_0^{2}=
\begin{pmatrix}
	0 & 0 & 0 \\
	0 & \taub^{\theta} & 0 \\
	0 & 0 & 0 \\
\end{pmatrix}\;,\\ 
\boldsymbol j_0^{\,3}&=
\begin{pmatrix}
	0 & 0 & 0 \\
	0 & 0 & 0 \\
	0 & 0 & \taub^{-\theta} \\
\end{pmatrix}\;.
\end{align*}
We have defined 
\begin{equation*}
\taub^{\theta}=e^{(i/4)\theta\tau_z}\taub e^{-(i/4)\theta\tau_z}\;,\quad
\taub=\tau_x \e_x +\tau_y \e_y\;,
\end{equation*}
where $\tau_{\nu}$ is the $\nu$-Pauli matrix. 
 
We need to compute 
\begin{equation*}
\underline{\boldsymbol j}^{\ell}(t)=
\begin{pmatrix}
	e^{i \mathcal H_{2}t} & 0 \\
	0  &  e^{i \mathcal H_{1}t} \\
\end{pmatrix}
\underline{\boldsymbol j}_0^{\ell}
\begin{pmatrix}
	e^{-i \mathcal H_{2}t} & 0 \\
	0  &  e^{-i \mathcal H_{1}t } \\
\end{pmatrix}
\end{equation*}
in the matrix representation $\vert L\rangle \langle L'\vert$, where each of $\vert L\rangle$ and $\vert L'\rangle$ is $(\vert b\rangle_1, \vert b\rangle_2, 0)^T$ or $(0, 0, \vert s\rangle )^T$ ($\ell=1,\,2,\,3$). 
For the initial $\mathfrak S$-transformed currents, we find
\begin{align}
	\underline{\boldsymbol j}_0^{1}&=\begin{pmatrix}
\tfrac{\alpha_{12}^2}{\lambda^2}\taub_1 & \tfrac{\alpha_{12}\alpha_{23}}{\lambda^2}\taub_c \\
\tfrac{\alpha_{12}\alpha_{23}}{\lambda^2}\taub_c^{\,\dag} & \left(\tfrac{\alpha_{23}}{\lambda}\right)^2\taub^{\,-\theta}\\	
\end{pmatrix}\\
&= \tfrac{1}{2}\begin{pmatrix}
\taub_1 & \taub_c \\
\taub_c^{\,\dag} & \taub^{\,-\theta}\\	
\end{pmatrix}\quad \mbox{if}\ \alpha_{12}=\alpha_{23}\;,\notag\\
\underline{\boldsymbol j}_0^{2}&=\begin{pmatrix}
\taub_2 & 0 \\
0 & 0\\	
\end{pmatrix}\;,\\
\underline{\boldsymbol j}_0^{3}&=\begin{pmatrix}
	\tfrac{\alpha_{23}^2}{\lambda^2}\taub_1 & -\tfrac{\alpha_{12}\alpha_{23}}{\lambda^2}\taub_c \\
	-\tfrac{\alpha_{12}\alpha_{23}}{\lambda^2}\taub_c^{\,\dag} & \tfrac{\alpha_{12}^2}{\lambda^2}\taub^{\,-\theta} \\
\end{pmatrix}\\
&= \tfrac{1}{2}\begin{pmatrix}
	\taub_1 & -\taub_c \\
	-\taub_c^{\,\dag} & \taub^{\,-\theta} \\
\end{pmatrix}\quad \mbox{if}\ \alpha_{12}=\alpha_{23}\;,  \notag 
\end{align}
where 
\begin{align*}
\taub_1&=
	\begin{pmatrix}
		\taub^{\,-\theta} & 0 \\
		 0   &   0 \\
	\end{pmatrix}\;, \quad 
	\taub_2=
	\begin{pmatrix}
		0 & 0 \\
		 0   &   \taub^{\,\theta} \\
	\end{pmatrix}\;,\notag\\
	\taub_c&=
	\begin{pmatrix}
		 \taub^{\,-\theta} \\
		 0 \\
	\end{pmatrix}\;.
\end{align*}
In fact, the $4\times 2$ matrix $\taub_c$ is responsible for $\bsigma_c$.

Thus, in the interaction picture we have
\begin{widetext}
\begin{align*}
	\underline{\boldsymbol j}^{1}(t)&=\frac{1}{\lambda^2}\begin{pmatrix}
		\alpha_{12}^2 e^{i \mathcal H_2 t} \taub_1 e^{-i \mathcal H_2 t} & \alpha_{12}\alpha_{23} e^{i\mathcal H_2 t} \taub_c e^{-i\mathcal H_1 t} \\
		\alpha_{12}\alpha_{23} e^{i\mathcal H_1 t} \taub_c^{\,\dag}e^{-i \mathcal H_2 t} & \alpha_{23}^2 e^{i\mathcal H_1 t} \taub^{\,-\theta} e^{-i\mathcal H_1 t}
	\end{pmatrix}\;,\quad  
\underline{\boldsymbol j}^{2}(t)=	
\begin{pmatrix}
	e^{i\mathcal H_2 t} \taub_2 e^{-i\mathcal H_2 t} & 0\\
	0 & 0\\
\end{pmatrix}\;,\\
\underline{\boldsymbol j}^{3}(t)&=\frac{1}{\lambda^2}
\begin{pmatrix}
		\alpha_{23}^2 e^{i\mathcal H_2 t} \taub_1 e^{-i \mathcal H_2 t} & -\alpha_{12}\alpha_{23} e^{i\mathcal H_2 t} \taub_c e^{-i\mathcal H_1 t} \\
		-\alpha_{12}\alpha_{23} e^{i\mathcal H_1 t} \taub_c^{\,\dag}e^{-i \mathcal H_2 t} & \alpha_{12}^2 e^{i\mathcal H_1 t} \taub^{\,-\theta} e^{-i\mathcal H_1 t}
	\end{pmatrix}\;.
	\end{align*}
\end{widetext}	
By abusing notation, we define the frequencies
\begin{align*}
\omega_{\eta\eta'}&=	(E_\eta-E_{\eta'})/\hbar\;,
\end{align*}
where $E_{\eta(\eta')}$ is the energy of either the effective single-layer or twisted bilayer system.
The frequency $\omega_{bs}$ mixes these energies, and will appear in $\bsigma_c$ (Sec.~\ref{subsec:cond-trilayer-details}).

Let us now outline the steps for the remaining computation of  $[\underline{\boldsymbol j}^{\ell}(t), \underline{\boldsymbol j}_0^{\ell'}]$. For every pair $(\ell, \ell')$ we use the matrices $\boldsymbol V_{\eta\eta'}^{\ell\ell'}$ from Eq.~\eqref{eq:V-tensor-def}, needed for $\eta=\eta'$. Note that the Hermitian adjoint of this expression reads
\begin{align*} 
\underline{\boldsymbol j}_0^{\ell'}\underline{\boldsymbol j}^{\ell}(t) &=\begin{pmatrix}
	\tvpb_{bb}^{\ell\ell'}(t)^\dagger  & \tvpb_{sb}^{\ell\ell'}(t)^\dagger \\
	\tvpb_{bs}^{\ell\ell'}(t)^\dagger & \tvpb_{ss}^{\ell\ell'}(t)^\dagger \\
\end{pmatrix}\;.
\end{align*}
Thus, it suffices to determine $\underline{\boldsymbol j}^{\ell}(t) \underline{\boldsymbol j}_0^{\ell'}$. The matrix elements of $[\underline{\boldsymbol j}^{\ell}(t), \underline{\boldsymbol j}_0^{\ell'}]$ are found from  $\tvpb^{\ell\ell'}_{\eta\eta}(t)-
(\tvpb^{\ell\ell'}_{\eta\eta}(t))^\dagger$.

\subsubsection{Tensor $\tvpb_{\eta\eta'}^{11}$}
\label{sssec:1-1}
For $(\ell, \ell')=(1,1)$, the diagonal block matrix of the first layer, $[\underline{\boldsymbol j}^{1}(t),\underline{\boldsymbol j}_0^{1}]$ directly comes from the following formulas. Regarding $\underline{\boldsymbol j}^{1}(t)\underline{\boldsymbol j}_0^{1}$, we compute
\begin{align*}
	\tvpb^{11}_{bb}&=\frac{\alpha_{12}^2}{\lambda^4}\left(\alpha_{12}^2e^{i\mathcal H_2 t} \taub_1 e^{-i\mathcal H_2 t} \taub_1 
	+\alpha_{23}^2e^{i\mathcal H_2 t} \taub_c \right.\\
	&\qquad \left. \times e^{-i \mathcal H_1 t} \taub_c^{\,\dag} \right)\;,
\end{align*}	
\begin{align*}
	\tvpb^{11}_{bs}&=\frac{\alpha_{12}\alpha_{23}}{\lambda^4}\left(\alpha_{12}^2 e^{i\mathcal H_2 t}\taub_1 e^{-i\mathcal H_2 t} \taub_c+ \alpha_{23}^2 e^{i\mathcal H_2 t}\taub_c \right. \\
	&\qquad \left. \times e^{-i\mathcal H_1 t} \taub^{-\theta}\right)\;,
\end{align*}
\begin{align*}	
	\tvpb^{11}_{sb}&=\frac{\alpha_{12}\alpha_{23}}{\lambda^4}\left(\alpha_{12}^2e^{i\mathcal H_1 t} \taub_c^{\,\dag} e^{-i\mathcal H_2 t} \taub_1 +\alpha_{23}^2 e^{i\mathcal H_1 t} \taub^{-\theta}\right. \\
	&\qquad \left. \times  e^{-i \mathcal H_1 t} \taub_c^{\,\dag} \right)\;,
\end{align*}
\begin{align*}	
	\tvpb^{11}_{ss}&= \frac{\alpha_{23}^2}{\lambda^4}\left(\alpha_{12}^2e^{i\mathcal H_1 t}\taub_c^{\,\dag} e^{-i\mathcal H_2 t} \taub_c+ \alpha_{23}^2e^{i\mathcal H t}\taub^{-\theta}\right. \\
	&\qquad \left. \times e^{-i\mathcal H_1 t} \taub^{-\theta}\right)\;.
\end{align*}

The product $\underline{\boldsymbol j}_0^{1}\underline{\boldsymbol j}^{1}(t)$ is thus determined by
\begin{align*}
	(\tvpb_{bb}^{11})^\dagger&=\frac{\alpha_{12}^2}{\lambda^4}\left(\alpha_{12}^2 \taub_1 e^{i\mathcal H_2 t} \taub_1 e^{-i\mathcal H_2 t} +\alpha_{23}^2 \taub_c \right.\\
	&\qquad \left. \times e^{i\mathcal H_1 t} \taub_c^{\,\dag} e^{-i \mathcal H_2 t} \right)\;,
\end{align*}
\begin{align*}	
	(\tvpb_{sb}^{11})^\dagger&=\frac{\alpha_{12}\alpha_{23}}{\lambda^4}\left(\alpha_{12}^2\taub_1 e^{i\mathcal H_2 t} \taub_c e^{-i\mathcal H_1 t} + \alpha_{23}^2\taub_c \right. \\
	& \qquad \left. \times e^{i\mathcal H_1 t} \taub^{-\theta} e^{-i\mathcal H_1 t} \right)\;,
\end{align*}
\begin{align*}	
	(\tvpb_{bs}^{11})^\dagger&= \frac{\alpha_{12}\alpha_{23}}{\lambda^4}\left(\alpha_{12}^2\taub_c^{\,\dag} e^{i\mathcal H_2 t} \taub_1 e^{-i\mathcal H_2 t}  +\alpha_{23}^2\taub^{-\theta}\right.\\
	&\qquad \left. \times  e^{i\mathcal H_1 t} \taub_c^{\,\dag}  e^{-i \mathcal H_2 t}  \right)\;,
\end{align*}
\begin{align*}	
	(\tvpb_{ss}^{11})^\dagger&= \frac{\alpha_{23}^2}{\lambda^4}\left(\alpha_{12}^2 \taub_c^{\,\dag} e^{i\mathcal H_2 t} \taub_c e^{-i\mathcal H_1 t} +\alpha_{23}^2\taub^{-\theta} \right. \\
	&\qquad \left. \times e^{i\mathcal H_1 t} \taub^{-\theta} e^{-i\mathcal H_1 t} \right)\;.
\end{align*}
The commutator $[\underline{\boldsymbol j}^{1}(t),\underline{\boldsymbol j}_0^{1}]$ is then readily computed.

\subsubsection{Tensors $\tvpb_{\eta\eta'}^{12}$ and $\tvpb_{\eta\eta'}^{21}$}
\label{sssec:2-1}
Let us now discuss the structure of both $[\underline{\boldsymbol j}^{1}(t),\underline{\boldsymbol j}_0^{2}]$ and $[\underline{\boldsymbol j}^{2}(t),\underline{\boldsymbol j}_0^{1}]$.
For the product $\underline{\boldsymbol j}^{1}(t)\underline{\boldsymbol j}_0^{2}$, we compute
\begin{align*}
	\tvpb^{12}_{bb}&= \frac{\alpha_{12}^2}{\lambda^2}e^{i\mathcal H_2 t} \taub_1 e^{-i\mathcal H_2 t} \taub_2\;,\notag\\
	\tvpb^{12}_{sb}&=\frac{\alpha_{12}\alpha_{23}}{\lambda^2} e^{i\mathcal H_1 t} \taub_c^{\,\dag} e^{-i\mathcal H_2 t} \taub_2\;,\notag\\
	\tvpb^{12}_{bs}&=0=\tvpb^{12}_{ss}\;.
\end{align*}
Similarly, for $\underline{\boldsymbol j}_0^{2}\underline{\boldsymbol j}^{1}(t)$ we invoke the Hermitian conjugate (adjoint) of each of the above quantities.

For $[\underline{\boldsymbol j}^{2}(t),\underline{\boldsymbol j}_0^{1}]$, a similar calculation yields
\begin{align*}
\tvpb_{bb}^{21}&=	\frac{\alpha_{12}^2}{\lambda^2}e^{i\mathcal H_2 t} \taub_2 e^{-i\mathcal H_2 t} \taub_1\;,\notag\\
\tvpb_{bs}^{21}&= \frac{\alpha_{12}\alpha_{23}}{\lambda^2} e^{i\mathcal H_2 t} \taub_2 e^{-i\mathcal H_2 t} \taub_c\;,\notag\\
\tvpb_{sb}^{21}&=0=\tvpb_{ss}^{21}\;.
\end{align*}
The Hermitian conjugate of each term follows directly. 

\subsubsection{Tensors $\tvpb_{\eta\eta'}^{13}$ and $\tvpb_{\eta\eta'}^{31}$}
\label{sssec:1-3}
Next, we focus on $[\underline{\boldsymbol j}^{1}(t),\underline{\boldsymbol j}_0^{3}]$ and $[\underline{\boldsymbol j}^{3}(t),\underline{\boldsymbol j}_0^{1}]$. For the former, the product $\underline{\boldsymbol j}^{1}(t)\underline{\boldsymbol j}_0^{3}$ is composed of
\begin{align*}
	\tvpb_{bb}^{13}&= \frac{\alpha_{12}^2\alpha_{23}^2}{\lambda^4}\left(e^{i\mathcal H_2 t} \taub_1 e^{-i\mathcal H_2 t} \taub_1 -e^{i\mathcal H_2 t} \taub_c e^{-i \mathcal H_1 t} \taub_c^{\,\dag} \right)\;,\notag\\
	\tvpb_{bs}^{13}&=\frac{\alpha_{12}^3\alpha_{23}}{\lambda^4}\left(e^{i\mathcal H_2 t}\taub_c e^{-i\mathcal H_1 t} \taub^{-\theta}- e^{i\mathcal H_2 t}\taub_1 e^{-i\mathcal H_2 t} \taub_c\right)\;,\notag\\
	\tvpb_{sb}^{13}&=\frac{\alpha_{12}\alpha_{23}^3}{\lambda^4}\left(e^{i\mathcal H_1 t} \taub_c^{\,\dag} e^{-i\mathcal H_2 t} \taub_1 -e^{i\mathcal H_1 t} \taub^{-\theta} e^{-i \mathcal H_1 t} \taub_c^{\,\dag} \right)\;,\notag\\
	\tvpb_{ss}^{13}&=\frac{\alpha_{12}^2\alpha_{23}^2}{\lambda^4}\left(e^{i\mathcal H_1 t}\taub^{-\theta} e^{-i\mathcal H_1 t} \taub^{-\theta}- e^{i\mathcal H_1 t}\taub_c^{\,\dag} e^{-i\mathcal H_2 t} \taub_c\right)\;.
\end{align*}
For $\underline{\boldsymbol j}_0^{3}\underline{\boldsymbol j}^{1}(t)$, we  compute the Hermitian adjoints.

On the other hand, in regard to  $[\underline{\boldsymbol j}^{3}(t),\underline{\boldsymbol j}_0^{1}]$ we obtain
\begin{align*}
	\tvpb_{bb}^{31}&= \frac{\alpha_{12}^2\alpha_{23}^2}{\lambda^4}\left(e^{i\mathcal H_2 t} \taub_1 e^{-i\mathcal H_2 t} \taub_1 -e^{i\mathcal H_2 t} \taub_c e^{-i \mathcal H_1 t} \taub_c^{\,\dag} \right)\;,\notag\\
	\tvpb_{bs}^{31}&=\frac{\alpha_{12}\alpha_{23}^3}{\lambda^4}\left(e^{i\mathcal H_2 t}\taub_1 e^{-i\mathcal H_2 t} \taub_c- e^{i\mathcal H_2 t}\taub_c e^{-i\mathcal H_1 t} \taub^{-\theta}\right)\;,\notag\\
	\tvpb_{sb}^{31}&=\frac{\alpha_{12}^3\alpha_{23}}{\lambda^4}\left(e^{i\mathcal H_1 t} \taub^{-\theta} e^{-i\mathcal H_1 t} \taub_c^{\,\dag} -e^{i\mathcal H_1 t} \taub_c^{\,\dag} e^{-i \mathcal H_2 t} \taub_1 \right)\;,\notag\\
	\tvpb_{ss}^{31}&=\frac{\alpha_{12}^2\alpha_{23}^2}{\lambda^4}\left(e^{i\mathcal H_1 t}\taub^{-\theta} e^{-i\mathcal H_1 t} \taub^{-\theta}- e^{i\mathcal H_1 t}\taub_c^{\,\dag} e^{-i\mathcal H_2 t} \taub_c\right)\;,
\end{align*}
along with their Hermitian conjugates. This step concludes the calculation for $(\ell,\,\ell')=(1,\,3),\,(3,\,1)$.

\subsubsection{Tensor $\tvpb_{\eta\eta'}^{22}$}
\label{sssec:2-2}
Consider the diagonal block matrix of the middle layer, related to $[\underline{\boldsymbol j}^{2}(t),\underline{\boldsymbol j}_0^{2}]$. For $\underline{\boldsymbol j}^{2}(t)\underline{\boldsymbol j}_0^{2}$, we need 
\begin{align*}
	\tvpb_{bb}^{22}&=e^{i\mathcal H_2 t} \taub_2 e^{-i\mathcal H_2 t} \taub_2\;,\notag\\
	\tvpb_{bs}^{22}&=0=\tvpb_{sb}^{22}=\tvpb_{ss}^{22}\;.
\end{align*}
The Hermitian adjoint, for  $\underline{\boldsymbol j}_0^{2}\underline{\boldsymbol j}^{2}(t)$, follows directly.

\subsubsection{Tensors $\tvpb_{\eta\eta'}^{23}$ and $\tvpb_{\eta\eta'}^{32}$}
\label{sssec:2-3}
For layers $2$ and $3$, we first compute
\begin{align*}
	\tvpb_{bb}^{23}&=\frac{\alpha_{23}^2}{\lambda^2}e^{i\mathcal H_2 t} \taub_2 e^{-i\mathcal H_2 t} \taub_1\;,\notag\\
	\tvpb_{bs}^{23}&=-\frac{\alpha_{12}\alpha_{23}}{\lambda^2} e^{i\mathcal H_2 t} \taub_2 e^{-i\mathcal H_2 t}\taub_c\;,\notag\\
	\tvpb_{sb}^{23}&=0=\tvpb_{ss}^{23}\;.
\end{align*}
Likewise, we have 
\begin{align*}
	\tvpb_{bb}^{32}&=\frac{\alpha_{23}^2}{\lambda^2}e^{i\mathcal H_2 t} \taub_1 e^{-i\mathcal H_2 t} \taub_2\;,\notag\\
	\tvpb_{sb}^{32}&=-\frac{\alpha_{12}\alpha_{23}}{\lambda^2} e^{i\mathcal H_1 t} \taub_c^{\,\dag} e^{-i\mathcal H_2 t} \taub_2\;,\notag\\
	\tvpb_{bs}^{32}&=0=\tvpb_{ss}^{32}\;.
\end{align*}

\subsubsection{Tensor $\tvpb_{\eta\eta'}^{33}$}
\label{sssec:3-3}
For the diagonal block matrix of the third (top) layer, a direct computation yields
\begin{align*}
	\tvpb^{33}_{bb}&=\frac{\alpha_{23}^2}{\lambda^4}\left(\alpha_{23}^2e^{i\mathcal H_2 t} \taub_1 e^{-i\mathcal H_2 t} \taub_1 +\alpha_{12}^2e^{i\mathcal H_2 t} \right.\\   & \left. \times  \taub_ce^{-i \mathcal H_1 t} \taub_c^{\,\dag} \right)\;,\notag\\
	\tvpb^{33}_{bs}&=-\frac{\alpha_{23}\alpha_{23}}{\lambda^4}\left(\alpha_{23}^2 e^{i\mathcal H_2 t}\taub_1 e^{-i\mathcal H_2 t} \taub_c+ \alpha_{12}^2 e^{i\mathcal H_2 t}\right.\\
	& \left. \times \taub_c e^{-i\mathcal H_1 t} \taub^{-\theta}\right)\;,\notag\\
	\tvpb^{33}_{sb}&=-\frac{\alpha_{23}\alpha_{23}}{\lambda^4}\left(\alpha_{23}^2e^{i\mathcal H_1 t} \taub_c^{\,\dag} e^{-i\mathcal H_1 t} \taub_1 +\alpha_{12}^2 e^{i\mathcal H_1 t} \right.\\
	& \left. \times \taub^{-\theta} e^{-i \mathcal H_1 t} \taub_c^{\,\dag} \right)\;,\notag\\
	\tvpb^{33}_{ss}&= \frac{\alpha_{12}^2}{\lambda^4}\left(\alpha_{23}^2e^{i\mathcal H_1 t}\taub_c^{\,\dag} e^{-i\mathcal H_2 t} \taub_c+ \alpha_{12}^2e^{i\mathcal H_1 t}\right.\\
	& \left. \times \taub^{-\theta} e^{-i\mathcal H_1 t} \taub^{-\theta}\right)\;.
\end{align*}

\subsection{Layer-resolved conductivity matrices}
\label{subsec:cond-trilayer-details}
Next, we outline the remaining steps of deriving the conductivity formulas of Sec.~\ref{subsec:main-res-trilayer} in this appendix. Let
\begin{align*}
 \vert s \rangle&= (\vert\zeta_{s1}\rangle, \vert\zeta_{s2}\rangle)^T,\\ 
 \vert b \rangle_1&= (\vert\varphi_{b1}\rangle, \vert\varphi_{b2}\rangle)^T\,,\quad  \vert b \rangle_2=(\vert\chi_{b1}\rangle, \vert \chi_{b2}\rangle)^T\;,
\end{align*}
where $\zeta_{si}(\bx)=\langle\bx\vert \zeta_{si}\rangle $, $\varphi_{bi}(\bx)=\langle \bx\vert \varphi_{bi}\rangle$ and $\chi_{bi}(\bx)=\langle\bx\vert \chi_{bi}\rangle$ ($i=1, 2$) are scalar Bloch wave functions. These are labeled by $s$ or $b$ which stand for the combined band index and quasi-momentum variable of the scaled Brillouin zone. The spin degeneracy factor, $g_s$, is omitted here but can be included in the end.

\subsubsection{$\bsigma^{11}$ tensor}
\label{sssec:11-cond-trilayer}
By Eq.~\eqref{eq:Kubo-1part-freq}, the structure of $\bsigma^{11}$ is dictated by the tensors $\tvpb^{11}_{\eta\eta}$. For example, for $\eta=b$ we have the term 
\begin{equation*}
\frac{\alpha_{12}^2}{\lambda^4}\left(\alpha_{12}^2e^{i\mathcal H_2 t} \taub_1 e^{-i\mathcal H_2 t} \taub_1 +\alpha_{23}^2e^{i\mathcal H_2 t} \taub_c e^{-i \mathcal H_1 t} \taub_c^{\,\dag} \right)\;.
\end{equation*}
For $\eta=s$, the quantities $\alpha_{12}$ and $\alpha_{23}$, $\mathcal H_{1}$ and $\mathcal H_{2}$, $\taub_{1}$ and $\taub^{\,-\theta}$, and $\taub_c$ and $\taub_c^{\,\dagger}$ must be respectively interchanged. By resolution of the identity, formula~\eqref{eq:sigma11-tri} emerges once we identify the $K$-valley contributions
\begin{align}
	(\sigma_{BL,\nu\nu'}^{11})_{K}&=-C_0\frac{e^2}{i\hbar} \frac{1}{\omega+i\delta}\sum_{bb'}(f_b-f_{b'})\notag\\
	&\qquad \times \frac{\langle b \vert \tau_{1\nu} \vert b'\rangle \langle b' \vert \tau_{1\nu'} \vert b\rangle}{\omega+\omega_{bb'}+i\delta}\;,\label{eq:sigmaBL11-def-app}
\end{align}
\begin{align}
	(\sigma_{SL,\nu\nu'})_{K}&=-C_0\frac{e^2}{i\hbar} 
	\frac{1}{\omega+i\delta}\sum_{ss'}(f_s-f_{s'}) \notag\\&\qquad \times \frac{\langle s \vert \tau^{-\theta}_\nu \vert s'\rangle \langle s' \vert \tau^{-\theta}_{\nu'} \vert s\rangle}{\omega+\omega_{ss'}+i\delta}\;,\label{eq:sigmaSL-def-app}
\end{align}
as $\delta\downarrow 0$ for real $\omega$.
Here, we set $\tau^{-\theta}_{(1)\nu}=\e_\nu\cdot \taub^{-\theta}_{(1)}$. These formulas lead to the forms $\bsigma_{SL}=\mathrm{diag}(\sigma_e^{(1)}, \sigma_e^{(1)})$ and $\bsigma_{BL}^{11}=\mathrm{diag}(\sigma_e^{(2)},\sigma_e^{(2)})$, with the suitable definitions of the scalars $\sigma_e^{(1,2)}$.
The coupling term $\bsigma_c$ comes from the $\taub_c^{(\dagger)}$-matrix terms in $\tvpb^{11}_{\eta\eta}$, viz. (with $\tau_{c\nu}=\e_\nu \cdot \taub_c$), 
\begin{align*}
(\sigma_{c,\nu\nu'})_K&=-C_0\frac{e^2}{i\hbar}
\frac{1}{\omega+i\delta}\sum_{bs}(f_b-f_s)\notag\\
&\times \left\{ \frac{\langle b \vert \tau_{c\nu} \vert s\rangle \langle s \vert \tau_{c\nu'}^\dagger \vert b\rangle}{\omega+\omega_{bs}+i\delta}  
- \frac{\langle b \vert \tau_{c\nu'} \vert s\rangle \langle s \vert \tau_{c\nu}^\dagger \vert b\rangle}{\omega-\omega_{bs}+i\delta}\right\}\;.
\end{align*}

We proceed to symmetrize the above quantities. For the effective single-layer system, by isotropy we have the diagonal element $\sigma_{SL,xx}=\sigma_{SL,yy}=\sigma_e^{(1)}$ with
\begin{align*}
\sigma_e^{(1)}(\omega)&=-i2C_0 g_v \frac{e^2}{\hbar(\omega+i\delta)} \sum_{ss'}(f_s-f_{s'})\notag\\
&\quad \times\frac{\omega_{ss'}}{(\omega+i\delta)^2-\omega_{ss'}^2}\,|\langle \zeta_{s1}\vert \zeta_{s'2}\rangle|^2\;.
\end{align*}
Here, $g_v$ is the valley degeneracy factor ($g_v=2$) and $\langle \zeta_{s1}\vert \zeta_{s'2}\rangle$ stands for the inner product of the scalar Bloch wave functions $\zeta_{1,2}^{s,s'}(\bx)$ over  the scaled cell. For the off-diagonal elements $\sigma_{SL, xy}$ and $\sigma_{SL, yx}$, symmetrization and the effect of isotropy yield 
\begin{equation*}
\sigma_{SL, xy}(\omega)=-\sigma_{SL, yx}(\omega)=0\;.
\end{equation*}

Similarly, for the effective twisted bilayer system, by isotropy we have $\sigma_{BL, xx}^{11}=\sigma_{BL,yy}^{11}=\sigma_e^{(2)}$ where
\begin{align*}
	\sigma_e^{(2)}(\omega)&=-i 2C_0 g_v \frac{e^2}{\hbar (\omega+i\delta)}\sum_{bb'}(f_b-f_{b'})\notag\\
	&\quad \times \frac{\omega_{bb'}}{(\omega+i\delta)^2-\omega_{bb'}^2} |\langle \varphi_{b1}\vert \varphi_{b'2}\rangle|^2\;.
\end{align*}
The calculation for $\sigma_{BL,xy}^{11}$ and $\sigma_{BL,yx}^{11}$ makes explicit use of time reversal symmetry and isotropy, yielding 
\begin{equation*}
\sigma_{BL,xy}^{11}(\omega)=-\sigma_{BL,yx}^{11}(\omega)=0\;.
\end{equation*}

Let us consider $\bsigma_c$. For its diagonal elements, we get
\begin{align*}
	(\sigma_{c,{xx \atop yy}})_K &=-2iC_0 \frac{e^2}{\hbar(\omega+i\delta)}\sum_{bs}(f_b-f_s)\notag\\
	& \times \frac{\omega_{bs}}{(\omega+i\delta)^2-\omega_{bs}^2}  |\langle \check\varphi_{b1}\vert \zeta_{s2}\rangle \pm \langle \varphi_{b2}\vert \check \zeta_{s1}\rangle |^2\;,	
\end{align*}
where $\check \varpi = e^{i\theta/2} \varpi$ for $\varpi=\varphi_{b1}, \zeta_{s1}$. By isotropy, we have $\sigma_{c,xx}(\omega)-\sigma_{c,yy}(\omega)=0$ which implies
\begin{align}\label{eq:sigma3-identity}
	&\sum_{bs}(f_b-f_s)\frac{\omega_{bs}}{(\omega+i\delta)^2-\omega_{bs}^2}\notag\\
	&\qquad \times \Rep(\langle \check\varphi_{b1}\vert\zeta_{s2}\rangle \langle \check\zeta_{s1}\vert\varphi_{b2}\rangle)=0\;;  
\end{align}
thus, $\sigma_{c,xx}=\sigma_{c,yy}=\sigma_{c}$ where
\begin{align}\label{eq:sigma3-tri-formula}
\sigma_{c}(\omega)&=-i 2C_0g_v \frac{e^2}{\hbar(\omega+i\delta)}\sum_{bs}(f_b-f_s)\frac{\omega_{bs}}{(\omega+i\delta)^2-\omega_{bs}^2}\notag\\
&\qquad \times \left(|\langle \varphi_{b1}\vert \zeta_{s2}\rangle|^2+|\langle \varphi_{b2}\vert \zeta_{s1}\rangle|^2\right)\;.	
\end{align}
This expression can be simplified by use of $|\langle \varphi_{b1}\vert \zeta_{s2}\rangle|= |\langle \zeta_{s1}\vert \varphi_{b2}\rangle|$ due to the invariance of $\mathcal H_{1,2}$ under the combined operations of sublattice index switch, complex conjugation, and parity inversion.

For the off-diagonal elements of $\bsigma_c$, we assert that
\begin{align*}
(\sigma_{c, {xy\atop yx}})_K&=2C_0\frac{e^2}{\hbar(\omega+i\delta)}\sum_{bs}(f_b-f_s)\notag\\
&\times \frac{\mp (\omega+i\delta) \Rep(m^{\varphi\zeta}_{bs})+i \omega_{bs}\Imp(m^{\varphi\zeta}_{bs})}{(\omega+i\delta)^2-\omega_{bs}^2},\ \delta\downarrow 0\;;
\end{align*}
\begin{equation*}
	m_{bs}^{\varphi\zeta}=(\langle\check\varphi_{b1}\vert \zeta_{s2}\rangle+\langle\varphi_{b2}\vert \check\zeta_{s1}\rangle)\ 
	(\langle\zeta_{s2}\vert \check\varphi_{b1}\rangle-\langle\check \zeta_{s1}\vert \varphi_{b2}\rangle)\;.
\end{equation*}
By symmetrizing this $K$-valley contribution, we obtain
\begin{align*}
	\sigma_{c,xy}(\omega)&=\sigma_{c,yx}(\omega)=i 2C_0g_v \frac{e^2}{\hbar(\omega+i\delta)}\notag\\&\times \sum_{bs}(f_b-f_s)\frac{\omega_{bs}}{(\omega+i\delta)^2-\omega_{bs}^2}\,\Imp(m_{bs}^{\varphi\zeta})\;.
\end{align*}
Notice that $\Imp(m_{bs}^{\varphi\zeta})=-2\Imp(\langle \check\varphi_{b1}\vert\zeta_{s2}\rangle \langle \check\zeta_{s1}\vert \varphi_{b2}\rangle)$. In view of identity~\eqref{eq:sigma3-identity}, along with the interchange																																																		 of the $\Rep(\cdot)$ and $\Imp(\cdot)$ factors in it, we infer that 
$\sigma_{c,xy}(\omega)=0=\sigma_{c,yx}(\omega)$. 
Alternatively, this vanishing of the off-diagonal elements of $\bsigma_c$ results from isotropy, by which $\sigma_{c,xy}(\omega)=-\sigma_{c,yx}(\omega)$.
Equations~\eqref{eq:sigma11-tri-diag-form} and~\eqref{eq:sigma0+} follow.

\subsubsection{$\bsigma^{12}$ and $\bsigma^{21}$ tensors}
\label{sssec:12-cond-trilayer}
Next, we address the cases of layer pairs $(\ell, \ell')=(1, 2)$, $(2, 1)$. Because each of $\tvpb^{12,21}_{bb}$ contains only one product with $\taub_{1,2}$ and is multiplied by $\alpha_{12}^2\lambda^{-2}$, while $\tvpb_{ss}^{12, 21}=0$, we see that each of $\bsigma^{12,21}$ is proportional to $\bsigma_{BL}^{12,21}$. For the $K$-valley term, Eq.~\eqref{eq:Kubo-1part-freq} implies 
\begin{align*}
	\bsigma^{12(21)}_{K}&=\frac{\alpha_{12}^2}{\lambda^2}(\bsigma_{BL}^{12(21)})_K\;,
\end{align*}
where
\begin{align*}
	(\sigma_{BL,\nu\nu'}^{12(21)})_{K}&=-C_0\frac{e^2}{\hbar (\omega+i\delta)}\sum_{bb'}(f_b-f_{b'})\notag\\
	&\times \frac{\langle b \vert \tau_{1(2)\nu} \vert b'\rangle \langle b' \vert \tau_{2(1)\nu'} \vert b\rangle}{i(\omega+i\delta+\omega_{bb'})}\;.
\end{align*}

For the diagonal elements, symmetrization entails
\begin{align*}
\sigma_{BL,{xx\atop yy}}^{12}&=\sigma_{BL,{xx\atop yy}}^{21}=-i 2C_0g_v \frac{e^2}{\hbar (\omega+i\delta)} \sum_{bb'}(f_b-f_{b'})\notag\\
&\times \frac{\omega_{bb'} \Rep\left\{\bigl(\langle \check\varphi_{b1}\vert \varphi_{b'2}\rangle  \pm \langle \varphi_{b2}\vert \check\varphi_{b'1}\rangle\bigr) \langle \check\chi_{b'2}\vert \chi_{b1}\rangle \right\}}{(\omega+i\delta)^2-\omega_{bb'}^2}\;.
\end{align*}
By isotropy, which means $\sigma_{BL,xx}^{12(21)}-\sigma_{BL,yy}^{12(21)}=0$, we have
\begin{equation}\label{eq:identity-sigma12-tri}
	\sum_{bb'}(f_b-f_{b'})\frac{\omega_{bb'}\Rep(\langle \check\varphi_{b1}\vert\varphi_{b'2}\rangle \langle \chi_{b'2}\vert\check\chi_{b2}\rangle)}{(\omega+i\delta)^2-\omega_{bb'}^2}=0\;;  
\end{equation}
therefore, we can set $\sigma_{BL,xx}^{12(21)}=\sigma_{BL,yy}^{12(21)}=\sigma_e^{12}$ where
\begin{align}\label{eq:sigma1-def-tri}
\sigma_e^{12}(\omega)&= -i 2C_0g_v \frac{e^2}{\hbar(\omega+i\delta)} \sum_{bb'}(f_b-f_{b'})\notag\\
&\quad  \times \frac{\omega_{bb'}\, \Rep\bigl(e^{-i\theta}\langle \varphi_{b1}\vert \varphi_{b'2}\rangle \langle \chi_{b'2}\vert \chi_{b1}\rangle)}{(\omega+i\delta)^2-\omega_{bb'}^2}  \;.
\end{align}

Regarding the off-diagonal elements of $\bsigma^{12}$, we find
\begin{align*}
\sigma_{BL,{xy\atop yx}}^{12}&=\sigma_{BL,{yx\atop xy}}^{21}=\pm i 2C_0g_v \frac{e^2}{\hbar(\omega+i\delta)} \sum_{bb'}(f_b-f_{b'})\notag\\
&\times \frac{\omega_{bb'}\Imp\left\{\bigl(\langle \check\chi_{b'2}\vert \chi_{b1}\rangle  \mp  \langle \chi_{b'1}\vert \check\chi_{b2}\rangle\bigr) \langle \check\varphi_{b1}\vert \varphi_{b'2}\rangle \right\}}{(\omega+i\delta)^2-\omega_{bb'}^2}\;.
\end{align*}
Note that $\sigma_{BL,{xy\atop yx}}^{12(21)}+\sigma_{BL,{yx\atop xy}}^{12(21)}=0$, inferred from Eq.~\eqref{eq:identity-sigma12-tri} by replacement of $\Rep(\cdot)$ by $\Imp(\cdot)$; thus, 
$\sigma_{BL,{xy\atop yx}}^{12(21)}=-\sigma_{BL,{yx\atop xy}}^{12(21)}=\sigma_{e,xy}^{12}$ with
\begin{align}\label{eq:chiral-tri-sigmac}
\sigma_{e,xy}^{12}(\omega)&=i 2C_0g_v \frac{e^2}{\hbar(\omega+i\delta)} \sum_{bb'}(f_b-f_{b'}) \notag\\
& \times \frac{\omega_{bb'}\, \Imp\bigl(e^{-i\theta}\langle \chi_{b'2}\vert \chi_{b1}\rangle  \langle \varphi_{b1}\vert \varphi_{b'2}\rangle\bigr)}{(\omega+i\delta)^2-\omega_{bb'}^2}\;.
\end{align}
Equations~\eqref{eq:sigma12-21-tri} and~\eqref{eq:sigma12-wo-mirror} follow.

\subsubsection{$\bsigma^{13}$ and $\bsigma^{31}$ tensors}
\label{sssec:13-cond-trilayer}
Consider  $(\ell,\ell')=(1,3)$ and $(3,1)$. By Eq.~\eqref{eq:Kubo-1part-freq}, $\bsigma^{13, 31}$ contain contributions from the two effective systems and the coupling term $\bsigma_c$ that are similar to those for  $\bsigma^{11}$. By inspection of  $\tvpb^{13,31}_{\eta\eta}$ in comparison to $\tvpb^{11}_{\eta\eta}$, from Eq.~\eqref{eq:Kubo-1part-freq} we can derive Eq.~\eqref{eq:sigma13-31-tri}. 

In more detail, for $\alpha_{12}=\alpha_{23}$, we see that 
\begin{equation*}
	\tvpb^{11}_{bb}=\frac{1}{4}\left(e^{i\mathcal H_2 t} \taub_1 e^{-i\mathcal H_2 t} \taub_1 +e^{i\mathcal H_2 t} \taub_c e^{-i \mathcal H_1 t} \taub_c^{\,\dag} \right)
\end{equation*}
whereas
\begin{equation*}
\tvpb_{bb}^{13}= \tfrac{1}{4}\left(e^{i\mathcal H_2 t} \taub_1 e^{-i\mathcal H_2 t} \taub_1 -e^{i\mathcal H_2 t} \taub_c e^{-i \mathcal H_1 t} \taub_c^{\,\dag} \right)=\tvpb_{bb}^{31}\;;
\end{equation*}
ditto for $\tvpb_{ss}^{11,13}$ via the switch of $\mathcal H_{2}$ and $\mathcal H_1$, $\taub_1$ and $\taub^{\,-\theta}$, and $\taub_c$ and its Hermitian adjoint. Thus, in the case with mirror symmetry, the matrix $\bsigma^{11}$ of Eq.~\eqref{eq:sigma11-m} is replaced by the $\bsigma^{13}$ given in Eq.~\eqref{eq:sigma13-31-m}. 

Now consider $\alpha_{12}\neq \alpha_{23}$. Notice the terms  in 
\begin{align*}
	\tvpb^{11}_{bb(ss)}&=\frac{\alpha_{12(23)}^2}{\lambda^4}\left(\alpha_{12(23)}^2e^{i\mathcal H_{2(1)} t} \taub_1^{(-\theta)} e^{-i\mathcal H_{2(1)} t} \taub_1^{(-\theta)} \right. \notag\\
	& \left. +\alpha_{23(12)}^2e^{i\mathcal H_{2(1)} t} \taub_c^{\,(\dagger)} e^{-i \mathcal H_{1(2)} t} \taub_{c}^{\,\dag (\,)} \right)\;,
\end{align*}
and map this to its 13- (and 31-) counterpart, viz.,
\begin{align*}
\tvpb_{bb(ss)}^{13}&= \frac{\alpha_{12}^2\alpha_{23}^2}{\lambda^4}\left(e^{i\mathcal H_{2(1)} t} \taub_1^{(-\theta)} e^{-i\mathcal H_{2(1)} t} \taub_1^{(-\theta)} \right. \notag\\
& \left. -e^{i\mathcal H_2 t} \taub_c^{\,(\dagger)} e^{-i \mathcal H_{1(2)} t} \taub_c^{\,\dag (\,)} \right). 	
\end{align*}
Equation~\eqref{eq:sigma13-31-tri} results from Eq.~\eqref{eq:sigma11-tri} via this map.

\subsubsection{$\bsigma^{22}$ tensor}
\label{sssec:22-cond-trilayer}
By inspection of $\tvpb^{22}_{\eta\eta}$, we notice that $\bsigma^{22}$ is identical to $\bsigma_{BL}^{22}$ according to the expression
\begin{align}
	(\sigma_{BL,\nu\nu'}^{22})_{K}&=-C_0\frac{e^2}{\hbar (\omega+i\delta)}\sum_{bb'}(f_b-f_{b'})\notag\\
	&\qquad \times \frac{\langle b \vert \tau_{2\nu} \vert b'\rangle \langle b' \vert \tau_{2\nu'} \vert b\rangle}{i(\omega+\omega_{bb'}+i\delta)}\;,\label{eq:sigmaBL22-def-app}
\end{align}
which comes from Eq.~\eqref{eq:sigmaBL11-def-app} after the replacement of $\taub_1$ by $\taub_2$. The resulting expression for $\bsigma^{22}=\bsigma_{BL}^{22}$ depends on $\alpha_{12, 23}$ through the effective Hamiltonian $\mathcal H_2$, and is a diagonal matrix (as is $\bsigma_{BL}^{11}$). Thus, we obtain Eq.~\eqref{eq:sigma22-tri}. The nonzero (diagonal) elements of $\bsigma_{BL}^{22}$ come from $\sigma_e^{(2)}$ by replacement of $\varphi_{1,2}$ by  $\chi_{1,2}$:
\begin{align*}
	\bsigma_{BL, xx}^{22}&=\bsigma_{BL,yy}^{22}=-i 2C_0g_v \frac{e^2}{\hbar (\omega+i\delta)}\sum_{bb'}(f_b-f_{b'})\notag\\
	&\quad \times \frac{\omega_{bb'}}{(\omega+i\delta)^2-\omega_{bb'}^2} |\langle \chi_{b1}\vert \chi_{b'2}\rangle|^2\notag\\
	&=\sigma_0^{22}=\sigma_e^{(2)}\;.
\end{align*}
In the above, we assumed that $|\langle \chi_{b1}\vert \chi_{b'2}\rangle|=|\langle \varphi_{b1}\vert \varphi_{b'2}\rangle|$, since $\mathcal H_{2,K}$ is invariant under the simultaneous operations of layer switch, complex conjugation, and $x$-direction reversal.

\subsubsection{$\bsigma^{23}$, $\bsigma^{32}$ and $\bsigma^{33}$ tensors}
\label{sssec:23-cond-trilayer}

Let us first consider $\bsigma^{23}$ and $\bsigma^{32}$. Evidently,  $\tvpb^{23,32}_{\eta\eta}$ come from $\tvpb^{21,12}_{\eta\eta}$ under the replacement of $\alpha_{12}$ with $\alpha_{23}$. By a rescaling and transposition of formula~\eqref{eq:sigma12-21-tri}, we obtain Eq.~\eqref{eq:sigma23-32-tri}. Recall that time reversal symmetry dictates $\bsigma_{BL}^{12}(\omega)=\bsigma_{BL}^{21}(\omega)^T$.

Regarding $\bsigma^{33}$, notice that $\tvpb^{33}_{\eta\eta}$ results from $\tvpb^{11}_{\eta\eta}$ by switch of the parameters $\alpha_{12}$ and $\alpha_{23}$. This implies that Eq.~\eqref{eq:sigma11-tri} yields Eq.~\eqref{eq:sigma33-tri}. Hence, matrix $\bsigma^{33}$ is diagonal. We set 
$\bsigma^{33}=\mathrm{diag}(\sigma_0^{33}, \sigma_0^{33})$ where the scalar $\sigma_0^{33}(\omega)$ directly comes from $\sigma_0^{11}(\omega)$ by the interchange of $\alpha_{12}$ and $\alpha_{23}$; cf. Eqs.~\eqref{eq:sigma0+} and~\eqref{eq:brevesigma0+}.



\section{In-plane currents as magnetization currents}
\label{app:magnetization}
 In this appendix, we show that the in-plane currents of any number $(n)$ of layers can always be written as the sum of the average current plus {\em magnetization} currents. The latter are imagined as coming from filling the space between adjacent layers with uniformly, in-plane  magnetized materials. The electric currents $(\J_i)$ are numbered from bottom to top, as in Fig.~\ref{fig:Geometry} for $n=3$. As in Sec.~\ref{subsec:in-plane-magn}, we use the symbol $\m_i $ to describe the magnetic moment (per unit surface) of the (imagined) magnetized material filling the space between layers $i$ and $i+1$. For the reference case of the trilayer system ($n=3$), we have three currents,  $\{\J_1, \J_2,\J_3 \} $, and two magnetized regions with magnetic moments per unit surface $\{\m_1, \m_2\}$,  filling the space between layers 1 and 2, and  layers 2 and 3, respectively.
 
 We can always write
\begin{equation}\label{eq:current-deviations}
\J_i = \frac{\J_{tot}}{n} + \j_i
,\end{equation} 
where $\J_{tot}=\sum_i \J_i$ is the total current, and $ \j_i$ represents the deviation from the average. Given that $ \sum_i \j_i = 0$,  each $ \j_i$ can be considered to be the sum of  the magnetization currents associated with the regions above and below the layer $i$. Therefore, we have:
\begin{equation}\label{eq:lower-js}
\frac{\m_{i-1} -\m_{i}}{l} = \e_z\times \j_{i}  
\, ,   \qquad  \qquad  1\leq i\leq n \, ,
\end{equation} 
with the constraint $ \sum_i \j_i = 0$, and $l$ is the distance between layers and $\e_z $  the normal unit vector. In~Eq.~\eqref{eq:lower-js}, it is understood that $\m_i=0 $ for $i$ outside its allowed range, that is, for $i\notin\{1, ...,n-1\}$.
Equation~\eqref{eq:lower-js}  provides the set of magnetizations $\{\m_i\} $ associated with any set of deviation currents $\{\j_i\} $, and vice versa. In the trilayer system, for instance, the use of Eqs.~\eqref{eq:current-deviations} and~\eqref{eq:lower-js},  together with  $ -i\omega\p = \J_{tot}$, leads to Eqs.~\eqref{moments:1}--\eqref{moments:3}  of Sec.~\ref{subsec:in-plane-magn}.

\bibliography{biblio}

\end{document}